\documentclass{aa} 

\usepackage{graphicx}
\usepackage{txfonts}
\usepackage{hyperref}

\newcommand{\teff}{\ensuremath{T_\mathrm{eff}}}
\newcommand{\logg}{\ensuremath{\log g}}
\newcommand{\kms}{$\rm km\,s ^{-1}$}
\newcommand{\vseni}{\ensuremath{v\sin{i}}}
\newcommand{\vrad}{\ensuremath{v_\mathrm{{rad}}}}
\newcommand{\Msol}{$\rm M_\odot$}

\begin{document} 

   \title{Young giants of intermediate mass}
   \subtitle{Evidence of rotation and mixing \thanks{Based on observations obtained
at Observatoire de Haute Provence, Canada-France-Hawaii
Telescope and Telescopio Nazionale Galileo
}}

   \author{Linda Lombardo
          \inst{1}
          \and
          Patrick Fran{\c c}ois \inst{1}
\and
Piercarlo Bonifacio \inst{1}
\and 
Elisabetta Caffau \inst{1}
\and
Aroa del Mar Matas Pinto \inst{1}
\and
Corinne Charbonnel \inst{2,3}
\and
Georges Meynet \inst{2}
\and
Lorenzo Monaco \inst{4}
\and
Gabriele Cescutti \inst{5,6,7}
\and 
Alessio Mucciarelli \inst{8,9}
          }

   \institute{GEPI, Observatoire de Paris, Universit\'e PSL, CNRS, 5 place Jules Janssen 92195 Meudon, France \\
              \email{Linda.Lombardo@observatoiredeparis.psl.eu}
         \and
             Department of Astronomy, University of Geneva, Chemin Pegasi 51, 1290, Versoix, Switzerland 
         \and
             IRAP, CNRS UMR 5277 \& Université de Toulouse, 14 Avenue Edouard Belin, 31400, Toulouse, France 
         \and
             Departamento de Ciencias Fisicas, Universidad Andres Bello, Fernandez Concha 700, Las Condes, Santiago, Chile 
         \and 
             INAF, Osservatorio Astronomico di Trieste, Via Tiepolo 11,  I-34143 Trieste, Italy 
         \and 
             IFPU, Istitute for the Fundamental Physics of the Universe, Via Beirut, 2, I-34151 Grignano, Trieste, Italy 
         \and
             INFN, Sezione di Trieste, Via A. Valerio 2, I-34127 Trieste, Italy 
         \and
             Dipartimento di Fisica e Astronomia, Università degli Studi di Bologna, Via Gobetti 93/2, 40129, Bologna, Italy 
         \and 
             INAF - Osservatorio di Astrofisica e Scienza dello Spazio di Bologna, Via Gobetti 93/3, 40129, Bologna, Italy 
             }

   \date{Received May 28, 2021; accepted September 24, 2021}

 
  \abstract
   {In the search of a sample of metal-poor bright giants using Str\"omgren photometry, we serendipitously found a sample of 26 young (ages younger than 1 Gyr) 
    metal-rich giants, some of which have high rotational velocities.}
   {We determined the chemical composition and rotational velocities of these stars
    in order to compare them with predictions from stellar evolution models. These stars
    where of spectral type A to B when on the main sequence,  and we therefore wished to compare their
    abundance pattern to that of main-sequence A and B stars.}
   {Stellar masses were derived by comparison of the position of the stars in the colour-magnitude
diagram with theoretical evolutionary tracks. These masses, together with Gaia photometry
and parallaxes, were used to derive the stellar parameters. We used spectrum synthesis and model atmospheres to determine  chemical abundances for 16 elements (C, N, O, Mg, Al, Ca, Fe, Sr, Y, Ba, La, Ce, Pr, Nd, Sm, and Eu) and
    rotational velocities. }
   {The age-metallicity degeneracy can affect photometric metallicity calibrations. 
    We identify 15 stars as likely binary stars. 
    All stars are in prograde motion around the Galactic centre and belong to the thin-disc population.
    All but one of the sample stars present low [C/Fe] and high [N/Fe] ratios together with constant [(C+N+O)/Fe], suggesting that they have undergone CNO processing and first dredge-up.
    The observed rotational velocities are in line with theoretical predictions of the evolution of rotating stars.     
      }
   {}

   \keywords{Stars: abundances --
                Stars: evolution --
                Stars: atmospheres
               }

   \maketitle
%

\section{Introduction}

The project Measuring at Intermediate metallicity Neutron Capture Elements (MINCE) (Cescutti et al. in preparation) has the goal of obtaining chemical abundances for stars in the
intermediate metallicity range ($\rm -2.5 \le [Fe/H] \le -1$).
 The aim is a detailed inventory
of the neutron-capture elements.

The target selection of the first few observational runs in the northern hemisphere heavily relied 
on Str\"omgren photometry. As detailed in Sect. \ref{selection}, this selection was unsuccessful in finding metal-poor giants. Its sample even proved to consist of young stars with masses in the range 2.5 to 6 solar masses in a narrow metallicity range
of about solar metallicity.

The most frequently studied G-K stars in this mass range are particular cases of peculiar stars, such as Ba stars \citep{BidelmanKeenan,Sneden,Antipova,Liang,AllenBarbuy,Smiljanic,Pereira,deCastro} and Cepheids \citep{Lemasle07,Lemasle08,Lemasle13,Genovali14,Genovali15}. 
As these stars were of A to B type when they were on the main sequence, the serendipitous discovery of this sample of giant stars allowed us to study this evolutionary stage directly. This stage is not very well characterised by observations so far because the time spent by stars in this phase is short.
In addition, it allows a direct comparison with the properties of A- to B-type stars.
For this reason, it is a unique opportunity for testing the predictions of stellar evolutionary models in terms of the evolution of chemical abundances and rotational velocities. 

We expect a large number of such stars to be observed in the 
course of wide-field surveys such as WEAVE \citep{Dalton} and 4MOST
\citep{4most}. The findings of our investigation can be used to select these stars from the wide surveys.

\section{Target selection}\label{selection}

   \begin{figure}
   \centering
\includegraphics[width=\hsize]{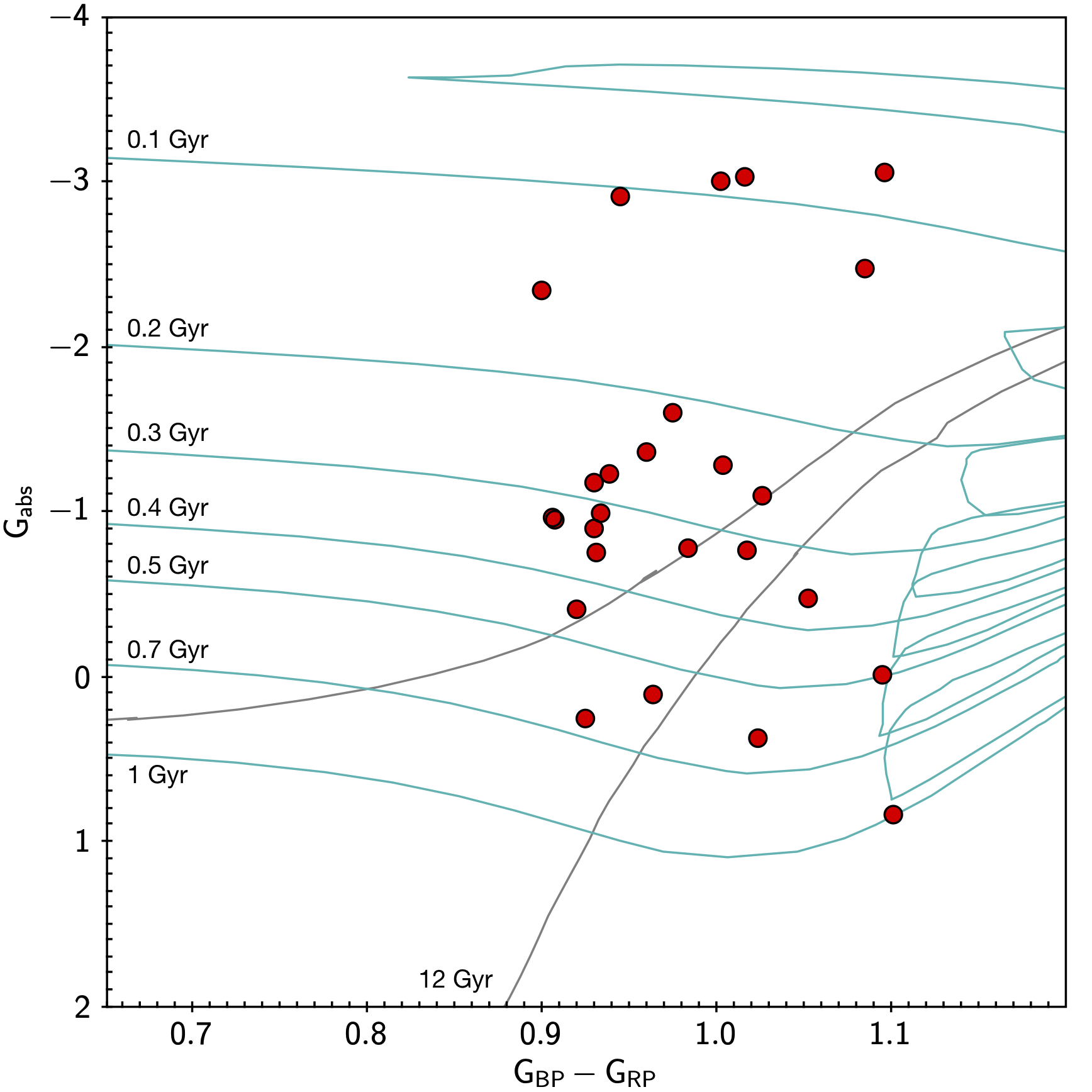}
\caption{Colour-magnitude diagram of Gaia EDR3 absolute magnitude $G$ and dereddened colour
$G_{BP}-G_{RP}$ comparing the observed stars with Parsec isochrones with solar metallicity and ages between 0.1 and 1 Gyr (cyan), and with [Fe/H]=--2.0 dex and age 12 Gyr (grey).}
         \label{Figdeg}
   \end{figure}

For the stars presented in this paper, we used the Str\"omgren photometry 
from the \citet{Paunzen} catalogue and the metallicity
calibration for giants
of \citet{Casagrande} to select candidate intermediate-metallicity stars.
To our surprise, all the
stars appeared to be at about solar metallicity when we inspected the spectra, 
and several rotate
rapidly (are therefore young and massive).

The question was whether a mistake was made in the star selection.
We examined the synthetic photometry of Parsec solar metallicity isochrones of ages in the range 0.1 to 1 Gyr \citep{Bressan}. 
As shown in Fig. \ref{Figdeg}, in the region of the Gaia Early Data Release 3 (EDR3) \citep{Gaia,GaiaEDR3} colour-magnitude diagram in which our targets lie, the young solar metallicity isochrones clearly overlap the locus occupied by the red giant branch (RGB) of old metal-poor isochrones. 
This is what is meant by age-metallicity degeneracy.
Furthermore, the Str\"omgren indices of many of the points along the young solar metallicity isochrones are within the validity range of the \citet{Casagrande} calibration. 
If they were used as input to the \citet{Casagrande} calibration, the majority of these photometric values would provide a metal-poor metallicity estimate.
This is another manifestation of the age-metallicity degeneracy.
The \citet{Casagrande} calibration in our opinion performs poorly on such metal-rich young stars because these stars were absent from the calibrators used by \citet{Casagrande} to define the calibration. 
However, even if these calibrators had been used, it is not clear if the degeneracy could have been lifted without introducing some age-sensitive quantity in the calibration.

\begin{table*}
\caption{Log of the observations.}
\label{tab:logobs}
\centering
\resizebox{\textwidth}{!}{
\begin{tabular}{lcclrrrrr}
\hline
\hline
Star &  date & MJD & Instrument & exposure time & \vrad & err \vrad & S/N at 680\,nm\\
     &       &     &            &      s        & \kms\ & \kms\ & \\
     \hline
     \\
  HD 191066      & 2019-09-13 & 58739.812 & SOPHIE@OHP1.93m & 344  & --9.350  & 0.002 & 60\\
  HD 192045      & 2019-09-13 & 58739.821 & SOPHIE@OHP1.93m & 3600 & --\,4.834  & 0.077 & 200\\
  HD 191066      & 2019-09-13 & 58739.867 & SOPHIE@OHP1.93m & 3600 & --9.348  & 0.001 & 230\\
  HD 205732      & 2019-09-13 & 58739.912 & SOPHIE@OHP1.93m & 3600 & --3.971  & 0.001 & 200\\
  HD 213036      & 2019-09-13 & 58739.956 & SOPHIE@OHP1.93m & 3600 & --\,40.89  & 0.128 & 200\\
  HD 217089      & 2019-09-14 & 58739.999 & SOPHIE@OHP1.93m & 3600 & --7.684  & 0.001 & 175\\
  HD 9637        & 2019-09-14 & 58740.043 & SOPHIE@OHP1.93m & 3600 & --2.361  & 0.002 & 210\\
  HD 21269       & 2019-09-14 & 58740.086 & SOPHIE@OHP1.93m & 1800 & --12.602 & 0.002 & 260\\
  HD 19267       & 2019-09-14 & 58740.11  & SOPHIE@OHP1.93m & 1800 &   2.697  & 0.001 & 170\\
  HD 13882       & 2019-09-14 & 58740.132 & SOPHIE@OHP1.93m & 3600 & --29.038 & 0.001 & 280\\
  HD 189879      & 2019-09-14 & 58740.878 & SOPHIE@OHP1.93m & 3600 & --29.532 & 0.001 & 200\\
  HD 195375      & 2019-09-14 & 58740.921 & SOPHIE@OHP1.93m & 3600 & --10.276 & 0.001 & 150\\
  HD 221232      & 2019-09-15 & 58740.965 & SOPHIE@OHP1.93m & 3600 & --30.649 & 0.001 & 170\\
  HD 219925      & 2019-09-15 & 58741.008 & SOPHIE@OHP1.93m & 3600 & --22.916 & 0.001 & 140\\
  HD 278         & 2019-09-15 & 58741.053 & SOPHIE@OHP1.93m & 3600 & --60.526 & 0.003 & 200\\
  HD 11519       & 2019-09-15 & 58741.141 & SOPHIE@OHP1.93m & 3183 & --11.260 & 0.001 & 120\\
  TYC2813-1979-1 & 2019-09-16 & 58742.101 & SOPHIE@OHP1.93m & 3600 & --17.163 & 0.002 & 60\\
  TYC2813-1979-1 & 2019-09-16 & 58742.144 & SOPHIE@OHP1.93m & 3295 & --17.177 & 0.002 & 40\\
  BD+42 3220     & 2019-11-20 & 58807.176 & ESPaDOnS@CFHT   & 2380 & --19.298 & 0.053 & 180\\
  BD+44 3114     & 2019-11-21 & 58808.176 & ESPaDOnS@CFHT   & 2380 & --15.216 & 0.055 & 260\\
  TYC 3136-878-1 & 2019-11-21 & 58808.205 & ESPaDOnS@CFHT   & 2380 &   1.034  & 0.053 & 280\\
  HD 40509       & 2019-12-21 & 58838.888 & HARPS-N@TNG     & 1800 & --1.656  & 0.005 & 280\\
  HD 41710       & 2019-12-21 & 58838.911 & HARPS-N@TNG     & 900  & --7.917  & 0.002 & 175\\
  HD 40655       & 2019-12-21 & 58838.928 & HARPS-N@TNG     & 2100 &   7.332  & 0.001 & 290\\
  HD 45879       & 2019-12-21 & 58838.955 & HARPS-N@TNG     & 1500 &   7.047  & 0.001 & 360\\
  HD 55077       & 2019-12-21 & 58838.974 & HARPS-N@TNG     & 1200 & --25.499 & 0.003 & 400\\
  HD 61107       & 2019-12-21 & 58838.991 & HARPS-N@TNG     & 1200 &   10.476 & 0.001 & 410\\
  HD 63856       & 2019-12-22 & 58839.007 & HARPS-N@TNG     & 1800 &   19.755 & 0.001 & 200\\

\hline
\end{tabular}
}
\end{table*}

\section{Observations}

The spectra used in this paper have been obtained
with three different telescopes and spectrographs.
The log of the observations and the observed radial velocities
are provided in Table\,\ref{tab:logobs}.

\subsection{SOPHIE at OHP}
SOPHIE \citep{SOPHIE} is a fibre-fed high-resolution spectrograph 
operated at the Observatoire de Haute Provence (OHP) 1.93\,m telescope.
The spectra were obtained in visitor mode during three
nights from September 13$^{th}$ to 16$^{th}$ 2019, the
observer was one of the co-authors (A.d.M. Matas Pinto).
The SOPHIE high-resolution mode, which provides
a resolving power $R\sim 75\,000$, was used for all
the observations. The spectral range we covered
is 387.2\,nm to 694.3\,nm. The wavelength calibration relied on a Th-Ar lamp and on a Fabry-P\'{e}rot {\it etalon}.
The data were reduced automatically  on the fly by
the SOPHIE pipeline.
Radial velocities (\vrad) were provided with the K5 template from the SOPHIE pipeline. For HD\,192045 and 
HD\,213036, the SOPHIE pipeline failed
because the input radial velocity, taken from the Gaia second data release 
\citep[hereafter Gaia DR2,][]{Gaia,GaiaDR2,Arenou,Sartoretti}, was too different
from the observed radial velocity of the star. The
\vrad\ were derived by us  
by measuring the cross-correlation function over the interval $\mathrm{500\  nm\leq\lambda\leq650\ nm}$.
A synthetic template with appropriate stellar parameters was used.

\subsection{ESPaDOnS at CFHT}
ESPaDOnS \citep{ESPADONS} is a
fibre-fed spectropolarimeter operated
at the 3.6\,m Canada-France-Hawaii telescope (CFHT)
on the summit of Mauna Kea.
The observation were obtained in the Queued Service
Observation mode of the CFHT in November 2019.
The spectroscopic mode ``Star+Sky'' was used, providing a resolving power of $R\sim 65\,000.$ It covers
the spectral range 370\,nm to 1051\,nm.
The data were delivered to us reduced
with the {\tt Upena} 
pipeline\footnote{\href{http://www.cfht.hawaii.edu/Instruments/Upena/}{http://www.cfht.hawaii.edu/Instruments/Upena/}}
, which uses the routines of the {\tt Libre-ESpRIT} software \citep{Donati}.
The output spectrum is provided in an order-by-order format.
We merged the orders using an ESO-MIDAS \footnote{\href{https://www.eso.org/sci/software/esomidas/}{https://www.eso.org/sci/software/esomidas/}} script written by ourselves. 
Radial velocities were derived measuring the cross-correlation function over the interval $\mathrm{420\  nm\leq\lambda\leq680\ nm}$.

\subsection{HARPS-N at TNG}

HARPS-N \citep{HARPS-N} is a fibre-fed high-resolution
spectrograph operated at the 3.5\,m Telescopio Nazionale Galileo (TNG) at the Canary
Island La Palma.
It is essentially a copy of HARPS \citep{HARPS}, operated by ESO on its 3.6\,m telescope
at La Silla.
The observations were obtained in service mode in December 2019. 
We used the high-resolution mode, 
which provides a resolving power $R\sim 115\,000$.
The wavelength range covered is 383\,nm to 690\,nm.
The data were reduced on the fly by the HARPS pipeline. 
Radial velocities were provided with the G2 template from the HARPS-N pipeline.

\section{Analysis}

\subsection{Stellar parameters} 

   \begin{figure}
   \centering
\includegraphics[width=\hsize]{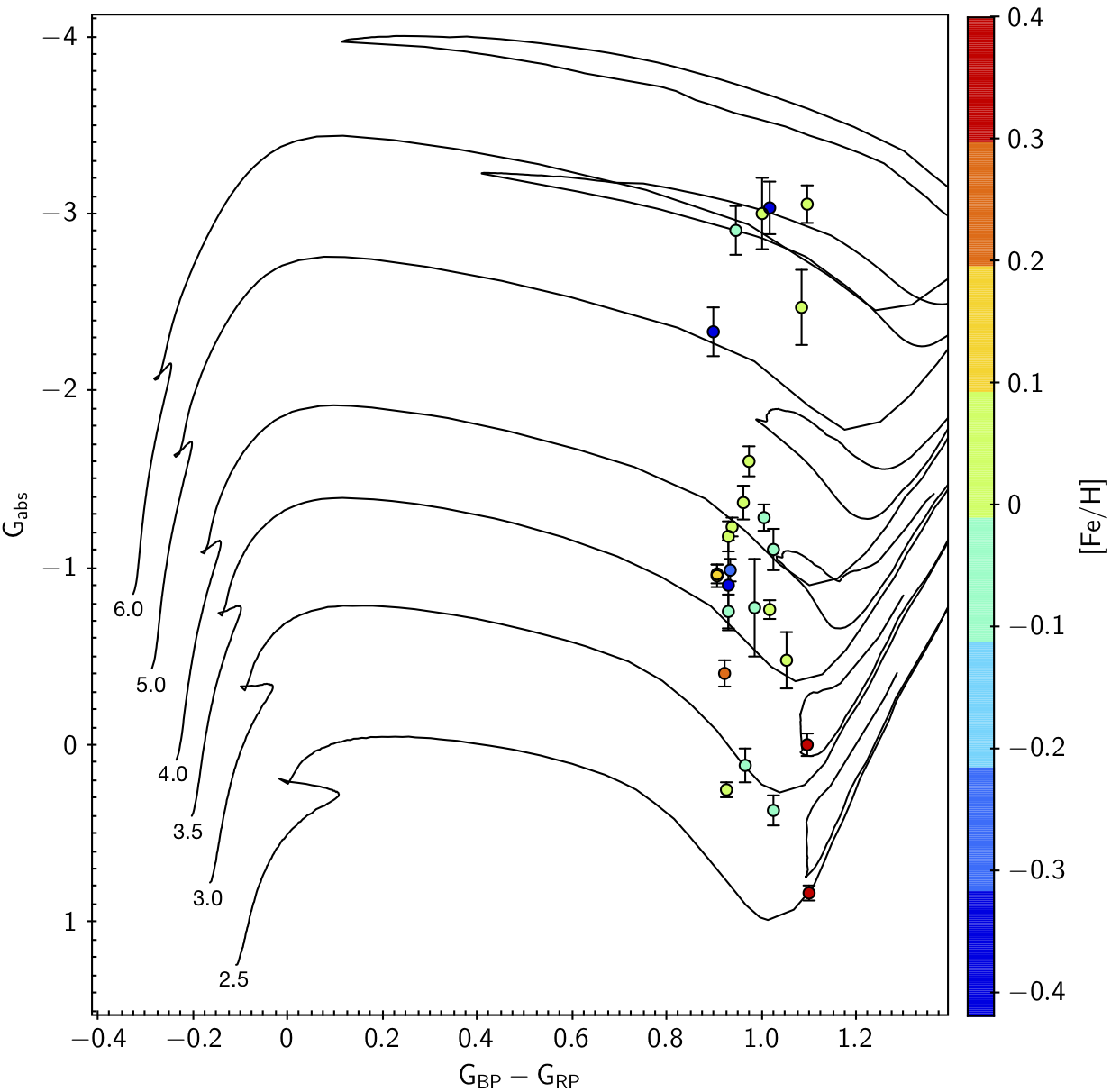}
\caption{Colour-magnitude diagram of Gaia EDR3 absolute magnitude  $G$ and dereddened colour
$G_{BP}-G_{RP}$ comparing the observed stars with \cite{Ekstrom} evolutionary tracks without rotation at metallicity Z\,=\,0.014 for stellar masses between 2.5 \Msol\ and 6.0 \Msol. Error bars represent the uncertainty on G$_{abs}$ , corresponding to 3$\sigma$ error on parallax. The colour index indicates our derived [Fe/H] of the stars.}
         \label{FigCmd}
   \end{figure}

\begin{table}
\caption{Range of atmospheric parameters of the ATLAS 9 model atmosphere grid.} 
\label{gridrange}
\centering
\begin{tabular}{l r r r}
\hline\hline
Parameter & Start & End & Step \\     
\hline
\teff & 3500 K & 5625 K & 125 K  \\
\logg & 0.00 dex & 3.00 dex & 0.50 dex \\
$[M/H]$ & --5.00 dex & --2.50 dex & 0.50 dex  \\
$[M/H]$ & --2.50 dex & 0.50 dex & 0.25 dex  \\
\hline
\end{tabular}
\end{table}

\begin{table*}
\caption{Coordinates, atmospheric parameters, metallicities, and rotational velocities of the observed stars.}
\label{tab:param}
\centering
\resizebox{\textwidth}{!}{
\begin{tabular}{rccrccccrrrrr}
\hline\hline 
Star & RA & DEC & G & Teff & logg & $\xi$ & Mass & {[}FeI/H{]} & {[}FeII/H{]} & \vseni & $\sigma$ & d \\
 & J2000 & J2000 & mag & K & {[}cgs{]} & \kms & \Msol & dex & dex & \kms &  &  \\
\hline
\\
HD 192045            & 20:12:18.44 & +16:13:01.6 & 7.068  & 5239 & 2.91 & 1.31 & 3.0 & --0.11 $\pm$ 0.16 & --0.27 $\pm$ 0.18 & 5.42  & 0.49 & 0.55 \\
HD 191066            & 20:07:28.92 & +17:01:13.9 & 7.411  & 5346 & 3.01 & 1.32 & 3.0 &   0.08 $\pm$ 0.20 & --0.18 $\pm$ 0.14 & 6.30  & 0.16 & 0.20 \\
HD 205732            & 21:36:08.54 & +41:55:57.1 & 7.560  & 5314 & 2.54 & 1.59 & 4.0 &   0.09 $\pm$ 0.16 & --0.22 $\pm$ 0.18 & 10.35 & 0.72 & 0.75 \\
HD 213036            & 22:27:33.13 & +51:58:35.9 & 7.556  & 5163 & 1.94 & 1.91 & 6.0 &   0.07 $\pm$ 0.16 & --0.19 $\pm$ 0.11 & 8.57  & 0.57 & 0.67 \\
HD 217089            & 22:57:26.91 & +46:49:21.6 & 7.825  & 5381 & 2.61 & 1.59 & 3.5 & --0.08 $\pm$ 0.13 & --0.05 $\pm$ 0.14 & 7.72  & 0.51 & 0.57 \\
HD   9637            & 01:35:19.49 & +44:55:34.6 & 8.476  & 5362 & 2.83 & 1.43 & 3.5 &   0.21 $\pm$ 0.10 &   0.29 $\pm$ 0.19 & 12.81 & 0.11 & 0.13 \\
HD  21269            & 03:27:02.31 & +34:25:46.2 & 6.386  & 4964 & 1.83 & 1.82 & 6.0 &   0.05 $\pm$ 0.26 &   0.14 $\pm$ 0.28 & 16.23 & 0.55 & 0.50 \\
HD  19267            & 03:09:22.16 & +64:17:58.8 & 6.550  & 5106 & 2.50 & 1.46 & 4.0 & --0.08 $\pm$ 0.12 &   0.21 $\pm$ 0.35 & 8.10  & 0.26 & 0.30 \\
HD  13882            & 02:16:49.26 & +51:31:44.2 & 7.179  & 5157 & 2.45 & 1.53 & 4.0 & --0.04 $\pm$ 0.14 &   0.24 $\pm$ 0.25 & 9.80  & 0.67 & 0.70 \\
HD 189879            & 20:00:12.51 & +47:55:57.7 & 8.017  & 4987 & 2.08 & 1.65 & 6.0 &   0.01 $\pm$ 0.17 & --0.02 $\pm$ 0.36 & 7.22  & 0.66 & 0.70 \\
HD 195375            & 20:29:08.30 & +45:44:04.5 & 7.878  & 5304 & 2.56 & 1.56 & 3.5 & --0.26 $\pm$ 0.13 &   0.10 $\pm$ 0.24 & 7.72  & 0.26 & 0.22 \\
HD 221232            & 23:29:53.55 & +50:46:38.9 & 8.183  & 5256 & 2.46 & 1.60 & 4.0 &   0.00 $\pm$ 0.12 &   0.25 $\pm$ 0.16 & 10.32 & 0.13 & 0.15 \\
HD 219925            & 23:19:10.68 & +49:13:59.9 & 8.539  & 5199 & 2.61 & 1.46 & 3.5 & --0.10 $\pm$ 0.17 & --0.35 $\pm$ 0.01 & 6.02  & 0.51 & 0.60 \\
HD    278            & 00:07:37.41 & +52:46:22.7 & 7.319  & 5133 & 2.59 & 1.42 & 3.5 &   0.06 $\pm$ 0.20 &   0.05 $\pm$ 0.20 & 22.17 & 0.57 & 0.60 \\
HD  11519            & 01:54:39.55 & +53:53:30.6 & 7.706  & 5230 & 2.35 & 1.65 & 4.0 &   0.04 $\pm$ 0.12 &   0.04 $\pm$ 0.08 & 6.87  & 0.46 & 0.50 \\
TYC 2813-1979-1      & 01:23:49.59 & +38:14:06.7 & 10.107 & 5058 & 2.67 & 1.32 & 3.5 &   0.08 $\pm$ 0.14 &   0.08 $\pm$ 0.05 & 6.37  & 0.45 & 0.55 \\
BD+42  3220          & 19:00:47.36 & +43:07:18.7 & 9.761  & 4982 & 3.02 & 1.09 & 2.5 &   0.40 $\pm$ 0.20 &   0.41 $\pm$ 0.07 & 4.00  & 0.14 & 0.10 \\
BD+44  3114          & 19:22:04.14 & +44:44:31.0 & 9.327  & 4990 & 2.77 & 1.22 & 3.0 &   0.37 $\pm$ 0.17 &   0.38 $\pm$ 0.10 & 5.98  & 0.26 & 0.30 \\
TYC 3136-878-1       & 19:46:28.83 & +39:15:59.4 & 10.602 & 5112 & 2.96 & 1.20 & 3.0 & --0.06 $\pm$ 0.16 & --0.07 $\pm$ 0.09 & 5.42  & 0.66 & 0.80 \\
HD  40509            & 06:03:48.81 & +64:05:20.1 & 7.731  & 5390 & 2.62 & 1.58 & 3.5 &   0.12 $\pm$ 0.18 &   0.32 $\pm$ 0.27 & 27.40 & 0.55 & 0.60 \\
HD  41710            & 06:07:46.85 & +22:42:27.6 & 7.104  & 5292 & 2.03 & 1.95 & 6.0 & --0.02 $\pm$ 0.13 &   0.17 $\pm$ 0.14 & 10.05 & 0.85 & 0.84 \\
HD  40655            & 06:00:55.34 & +17:53:24.8 & 7.844  & 5379 & 2.21 & 1.88 & 5.0 & --0.35 $\pm$ 0.11 & --0.13 $\pm$ 0.13 & 9.52  & 1.08 & 1.20 \\
HD  45879            & 06:30:57.65 & +21:48:17.0 & 7.232  & 5112 & 1.90 & 1.90 & 6.0 & --0.36 $\pm$ 0.16 &   0.07 $\pm$ 0.08 & 11.81 & 0.88 & 1.05 \\
HD  55077            & 07:16:01.93 & +64:46:51.6 & 7.077  & 5297 & 2.59 & 1.54 & 3.5 & --0.42 $\pm$ 0.11 & --0.25 $\pm$ 0.23 & 22.96 & 0.74 & 0.77 \\
HD  61107            & 07:39:25.81 & +40:40:16.8 & 7.137  & 5332 & 2.56 & 1.58 & 4.0 &   0.05 $\pm$ 0.15 &   0.17 $\pm$ 0.21 & 15.30 & 0.80 & 0.75 \\
HD  63856            & 07:51:25.12 & +11:09:56.5 & 7.888  & 5321 & 2.67 & 1.50 & 3.5 & --0.05 $\pm$ 0.16 & --0.32 $\pm$ 0.08 & 5.32  & 0.46 & 0.56 \\
\hline 
\end{tabular}
}
\tablefoot{Uncertainties on \vseni\ are expressed in $\sigma$ and in mid-range (d).}
\end{table*}

To derive stellar parameters, we used Gaia EDR3 photometry ($G$ and $G_{BP}-G_{RP}$) and parallaxes.
We defined a grid in the parameter space using the ATLAS 9 model atmosphere grids by Mucciarelli et al. (in prep.). The range of atmospheric parameters covered by the grid is shown in Table \ref{gridrange}.

We computed theoretical values of $G_{BP}-G_{RP}$, bolometric correction ($BC_G$), and extinction coefficients $A_{G}$, E($G_{BP}-G_{RP}$) using the reddening law of \citet{Fitzpatrick} for the whole grid.
Effective temperatures (\teff) and surface gravities (\logg) were derived iteratively with the following procedure:

\begin{enumerate}

\item The mass and the metallicity of the star were fixed at input values.

\item \teff\ was derived by interpolating in $G_{BP}-G_{RP}$ at fixed metallicity, and the bolometric correction was derived by interpolation from the new \teff.

\item \logg\ was derived using \teff\ and bolometric correction found in the previous step from the equation
\begin{equation}\label{eqlogg}
\begin{split}
   \logg = & \log (M/M_\odot) + 4\log (\teff/T_\odot) +0.4\ (G_0 + BC_{G}) \\
   & +2\log p + \log L_\odot + \logg_\odot
\end{split}    
\end{equation}
where $M$ is the stellar mass, $G_0$ is the dereddened apparent G magnitude, $BC_G$ is the bolometric correction, and $p$ is the parallax. 
\item $A_{G}$ and E($G_{BP}-G_{RP}$) were derived by interpolating in the theoretical grid adopting the E(B-V) from STILISM maps \citep{STILISM}.

\item $G$ and $G_{BP}-G_{RP}$ were dereddened using E($G_{BP}-G_{RP}$). 

\item The procedure was iterated until the difference between the new and the old \teff \ was smaller than $\pm$ 50 K and the difference between the new and the old \logg\ was smaller than 0.05 dex.  

\end{enumerate}

After we derived the stellar parameters, we used the measured equivalent widths (EWs)
of the \ion{Fe}{i} lines (see Sect\,\ref{abu})\ and
the GALA code \citep{GALA} to derive the metallicity of the stars. With the new metallicity values, we again derived the stellar parameters and stopped the iteration when the difference between the new and the old \teff \ was smaller than $\pm$ 50 K.
To confirm the values of \teff\ and \logg\ obtained from the procedure, we derived effective temperatures and surface gravities using a new implementation of the \citet{Mucciarelli&Bellazzini} InfraRed Flux Method (IRFM) colour-\teff\ calibration for giants stars based on EDR3 data \citep{Mucciarelli2021}. 
The two effective temperatures agree well, in particular, \teff\ from IRFM calibration is 70\,K cooler on average than those estimated with the method described above.

As final parameters, we adopted \teff\ and \logg\ derived from the \citeauthor{Mucciarelli2021} calibration because the IRFM method is less dependent on the adopted models with respect to the method described in the procedure. We fixed the uncertainty on \teff\ at $\mathrm{\pm 100\ K}$, according to the dispersion of the \citeauthor{Mucciarelli2021} calibration. 
In Table \ref{tab:param} we present the stellar parameters.
Microturbulent velocities ($\xi$) were estimated using the calibration derived by \cite{DutraFerreira}. The uncertainty on $\xi$ is $\pm 0.1$ \kms\  according to the uncertainty of \citeauthor{DutraFerreira} calibration.

In Fig.\,\ref{FigCmd} we compare Gaia EDR3 photometry of the observed stars to the \cite{Ekstrom} evolutionary tracks without rotation at solar metallicity for stellar masses between 2.5 \Msol\ and 6.0 \Msol\ provided by the SYCLIST code \citep{Syclist}, or interpolated between fully computed tracks. From Fig.\,\ref{FigCmd} we deduce that these stars are very young, with ages between 0.1 Gyr (0.06 Gyr if we consider the evolutionary track for 6 \Msol) and 0.55 Gyr. The stellar masses we deduced from the evolutionary tracks are listed in Table\,\ref{tab:param}. 
The uncertainty on stellar mass depends on the parallax error and on the stellar model adopted. If we adopt stellar models with rotation, for example, the evolutionary tracks are more luminous than the respective tracks without rotation (see Fig. 10 in \citealt{Georgy}). For this reason, we estimate an uncertainty in mass from 0.5 \Msol\ for the less massive stars to 1 \Msol\ for those that lie in the region of the colour-magnitude diagram in which different evolutionary tracks overlap. We estimated an uncertainty of $\pm 0.1$ dex on \logg\ by varying the stellar mass in Eq. \ref{eqlogg} according to the mass error.

\subsection{Iron abundances}\label{abu}

\begin{table*}
\caption{List of \ion{Fe}{I} lines used to derive the rotational velocity of stars.}
\label{tab:3}
\centering
\resizebox{\textwidth}{!}{
\begin{tabular}{rcccccccccccc}
\hline\hline 
 & 5778\AA & 5809\AA & 6096\AA & 6151\AA & 6380\AA & 6627\AA & 6726\AA & 6752\AA & 6806\AA & 6810\AA & 6820\AA & 6862\AA \\
 \hline
 \\
HD 192045 &  &  & x & x & x &  &  &  & x &  &  &  \\
HD 191066 &  &  & x & x &  &  & x &  & x &  &  &  \\
HD 205732 &  &  &  & x &  &  &  &  & x & x & x &  \\
HD 213036 & x & x & x &  &  &  &  &  &  &  & x &  \\
HD 217089 &  &  &  & x & x &  &  & x & x &  &  &  \\
HD   9637 &  &  &  & x & x &  &  & x &  & x &  &  \\
HD  21269 &  &  &  & x &  &  &  & x & x &  &  &  \\
HD  19267 &  &  &  &  &  &  & x &  & x & x & x &  \\
HD  13882 &  &  &  &  & x &  &  & x &  & x & x &  \\
HD 189879 &  &  & x &  &  &  & x &  & x &  & x &  \\
HD 195375 &  &  &  & x & x &  &  & x & x &  &  &  \\
HD 221232 &  &  &  & x & x &  &  & x &  &  & x &  \\
HD 219925 &  &  & x & x &  &  & x &  & x &  &  &  \\
HD    278 &  &  &  & x & x &  &  & x & x &  &  &  \\
HD  11519 &  &  &  &  &  &  & x & x & x &  & x &  \\
TYC 2813-1979-1 &  &  & x &  &  &  & x & x &  &  & x &  \\
BD+42  3220 &  &  &  & x &  &  &  &  &  & x &  &  \\
BD+44  3114 &  &  &  &  &  & x & x &  &  & x &  & x \\
TYC 3136-878-1 &  &  &  & x & x &  &  &  & x &  & x &  \\
HD  40509 &  &  &  & x & x &  &  & x & x &  &  &  \\
HD  41710 &  &  &  & x & x &  &  & x & x &  &  &  \\
HD  40655 &  &  &  & x & x &  &  & x & x &  &  & \\
HD  45879 &  &  &  & x & x &  &  & x & x &  &  &  \\
HD  55077 &  &  &  & x & x &  &  & x & x &  &  &  \\
HD  61107 &  &  &  & x & x &  &  & x & x &  &  &  \\
HD  63856 &  & x & x & x &  &  &  &  & x &  &  &  \\
\hline 
\end{tabular}%
}
\tablefoot{The X represents the lines we used for each star.}
\end{table*}

Fe abundances were derived from the EW of the lines.  
The EWs were measured with the FITLINE code developed by P. François \citep{Lemasle07}.
FITLINE is a semi-interactive FORTRAN program that measures EWs of high-resolution spectra using genetic algorithms \citep{Charbonneau}. 
Lines are fitted by a Gaussian defined by four parameters: the central wavelength, the width and depth of the line, and the continuum value.
For each line, the algorithm runs as follows: 
1) the program generates an initial set of Gaussians, giving random values to the four parameters of the Gaussian.
2) The fit quality is estimated by calculating the $\chi^2$.
3) A new ”generation” of Gaussians is calculated from the 20 best fits after adding random modifications to the initial set of parameters.
4) The new set of parameters replaces the old set, and its accuracy is estimated again using a $\chi^2$ evaluation.
Finally, 5) the process is iterated until the convergence to the best Gaussian fit is achieved.
For the most highly rotating stars (HD 21269, HD 278, HD 40509, HD 55077, and HD 61107) FITLINE was modified to take the Gaussian and rotational profile of the lines into account. 
To derive Fe abundances from EWs, we used the GALA code \citep{GALA}, which compares the measured EW for each line with the theoretical EW computed from the curve of growth of the line. 
\ion{The Fe}{I} and \ion{Fe}{II} abundances we derived for this sample of stars are listed in Table \ref{tab:param}.

\subsection{Rotational velocities}

\begin{figure}
\centering
\includegraphics[width=\hsize]{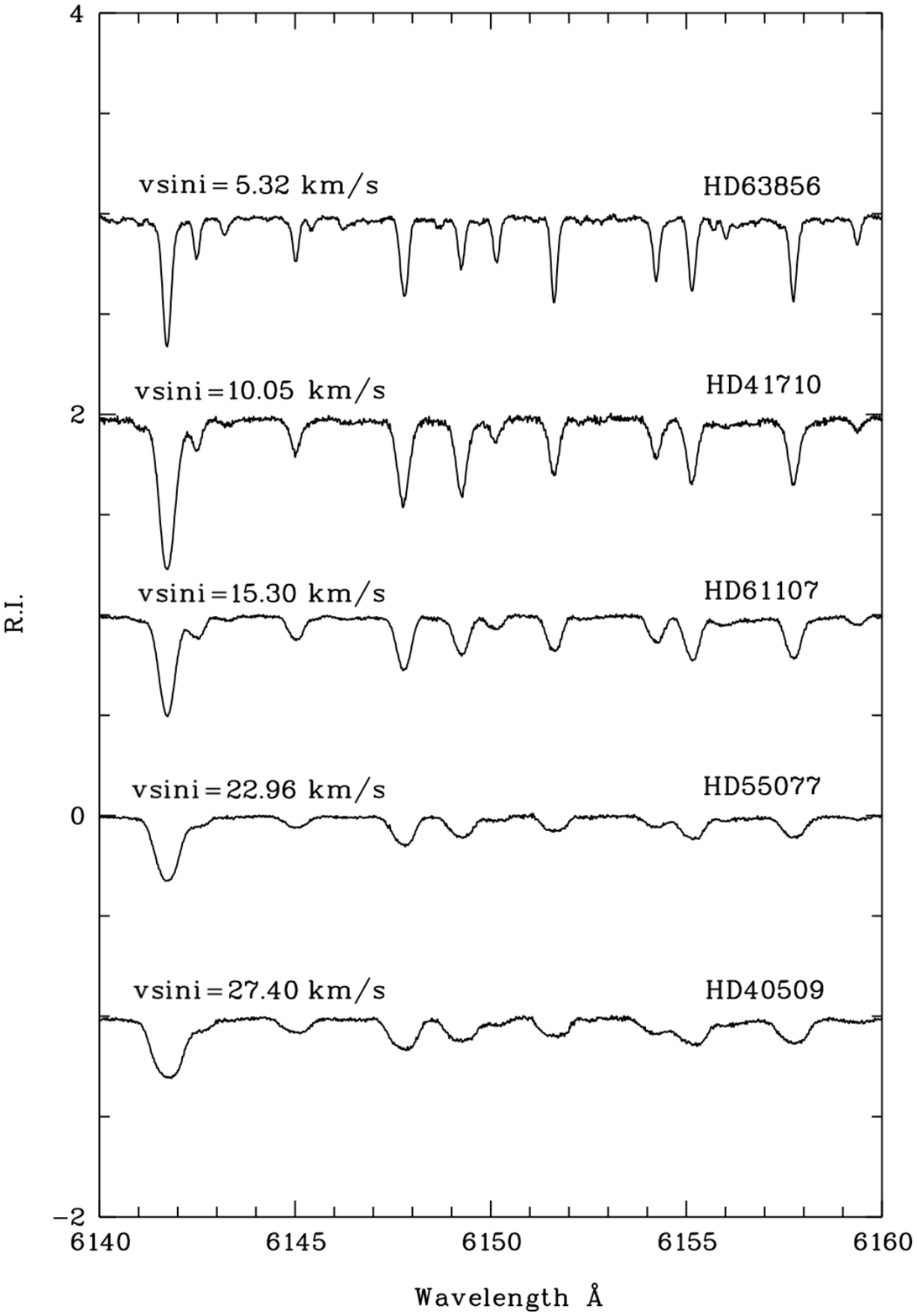}
\caption{Spectra of five stars with similar \teff\ and different \vseni. The spectra have been normalised and are shifted to facilitate visualisation. }
\label{spectra}
\end{figure}

Rotational velocities (\vseni) were measured by fitting the observed line profiles with the corresponding theoretical line profiles that were obtained from a synthetic spectrum computed with the spectral synthesis code SYNTHE \citep{Kurucz, SYNTHE} based on ATLAS\,9 1D plane-parallel model atmospheres \citep{Kurucz}. In order to better compare \vseni\ between different stars, we selected a set of \ion{Fe}{I} lines to be fitted. 
The list of \ion{Fe}{I} lines we used to measure \vseni\ is shown in Table \ref{tab:3}. 
For each line, we performed a $\chi^2$ minimisation fit on the observed line profile using three synthetic spectra computed with different rotational velocities. The metallicity of the synthetic spectra was fixed at the value of iron abundance derived from the EW of the single line. We made the assumption that the line broadening of synthetic spectra was equal to the instrumental broadening. 
The values of \vseni\ we obtained are presented in Table \ref{tab:param}. 
This approach does not allow us to distinguish between \vseni\ and other sources of line broadening, such as macroturbulence. The spectra of five stars with similar \teff\ but different \vseni\ are shown in Fig. \ref{spectra}. For the faster-rotating stars, the line profile is dominated by the rotational profile, and we can assume that other broadening effects are negligible with respect to \vseni. 
For the more slowly rotating stars, however, the contribution of macroturbulence in line broadening is comparable to the rotational contribution, so that the values of \vseni\ obtained for these stars must be interpreted as upper limits.

\subsection{Chemical abundances of other elements}
We were able to derive for the stars in the sample the elemental abundances of C, N, O, Mg, Al, and Ca and for the neutron capture elements Sr, Y, Ba, La, Ce, Pr, Nd, Sm, and Eu. For all elements we adopted the solar abundances determined by \cite{Caffau} and \cite{Lodders} (see Table \ref{solar}). The chemical abundances we obtained are listed in Tables\,\ref{tab:cno}, \ref{tab:mgalca}, and \ref{tab:ncap} in the appendix. The values A(X) are expressed in the form $A(X) = \log(X/H) + 12 $. The abundance ratios 
[X/Fe]\footnote{[X/Fe] = $\log_{10}(X/Fe) - \log_{10}(X/Fe)_\odot$.} 
are expressed as $\mathrm{[X/Fe] = [X/H] - [Fe/H]}$ for elements up to Ca and as $\mathrm{[X/Fe] = [X/H] - [FeII/H]}$ for O and n-capture elements.

The carbon abundance was derived from the G-band by minimisation of the $\chi ^2$ when we compared the observed spectrum to a grid of synthetic spectra with different C abundances.
The synthetic spectra were computed with SYNTHE \citep{Kurucz} using the ATLAS\,9 model \citep{Kurucz} computed for each star  \citep[see][for details]{abbo}.
The G-band is strong for all the stars, and we estimate an uncertainty in the range 0.2 to 0.5\,dex in the C abundances, which is mainly related to the continuum placement.
The nitrogen abundances were determined in a similar way using
the violet  CN band at 412.5\,nm, assuming the C abundance
derived from the G-band to be fixed. The error is mainly due to the uncertainty
of placing the continuum and is in the range 0.3 to 0.6\, dex.
Oxygen was determined from the EW of the [OI]
630\,nm line. This was measured with the 
iraf\footnote{https://iraf-community.github.io/} task {\tt splot} only when the 
line was not affected by blends with telluric lines or
the blend was minor and could be taken into account using the deblend option of {\tt splot}. 
Mg, Al, and Ca abundances were derived using the procedure described in section \ref{abu}. 
The uncertainty showed in Table \ref{tab:mgalca} represents the line-to-line scatter if the abundance was derived from $\geq$\,2 lines, otherwise it represents the abundance error due to continuum placement. 
For the neutron-capture elements, the abundance was determined by matching the observed spectrum around each line of the list with a synthetic spectrum computed using the local thermodynamic equilibrium (LTE) spectral line analysis code turbospectrum \citep{alvarez1998,plez2012}, which treats scattering in detail. 
The computed errors in the elemental abundances ratios due to uncertainties in the stellar parameters are listed in Table \ref{errors}. The errors were estimated varying  \teff\ by $\pm$ 100~K, \logg\  by $\pm$ 0.5~dex, and $\xi$  by $\pm$ 0.5 dex in the model atmosphere of HD 13882. The results for other stars are similar. The main uncertainty comes from the error in the continuum placement when the synthetic line profiles are matched to the observed spectra. This error is of the order of 0.1 dex.  When several lines are available, the typical line-to-line scatter for a given element is 0.1 dex.

\begin{table}[ht]
\caption{Solar abundance values adopted in this work. } 
\label{solar}
\centering
\begin{tabular}{l c c }
\hline\hline
Element & A(X) & References \\     
\hline
C & 8.50 & \cite{Caffau}  \\
N & 7.86 & \cite{Caffau}  \\
O & 8.76 & \cite{Caffau}  \\
Mg & 7.54 & \cite{Lodders} \\ 
Al & 6.47 & \cite{Lodders} \\ 
Ca & 6.33 & \cite{Lodders} \\ 
Fe & 7.52 & \cite{Caffau} \\ 
Sr & 2.92 & \cite{Lodders} \\
Y  & 2.21 & \cite{Lodders} \\
Ba & 2.17 & \cite{Lodders} \\
La & 1.14 & \cite{Lodders} \\  
Ce & 1.61 & \cite{Lodders} \\
Pr & 0.76 & \cite{Lodders} \\
Nd & 1.45 & \cite{Lodders} \\
Sm & 1.00 & \cite{Lodders} \\
Eu & 0.52 & \cite{Caffau} \\
\hline
\end{tabular}
\end{table}

\section{Possible binary stars}

\begin{table}[ht]
\caption{Comparison between Gaia DR2 and observed radial velocities for our stars.}
\label{tab:gaiarv}
\centering
\resizebox{\hsize}{!}{
\begin{tabular}{lrrc}
\hline\hline 
Star & \vrad$_{GDR2}$ & \vrad  & flag \\
 & \kms & \kms & \\
\hline
HD 192045           & --\,43.846 $\pm$ 2.253 &  --\,4.834  $\pm$ 0.077 & $\bullet$           $\circ$           $\star$ \\ 
HD 191066           &    --5.911 $\pm$ 3.451 &    --9.349  $\pm$ 0.001 & $\bullet$           \phantom{$\circ$} $\star$ \\ 
HD 205732           &    --3.150 $\pm$ 0.147 &    --3.971  $\pm$ 0.001 & \phantom{$\bullet$} $\circ$           \phantom{$\star$}\\ 
HD 213036           &  --\,4.369 $\pm$ 1.662 & --\,40.890  $\pm$ 0.128 & $\bullet$           $\circ$           $\star$ \\ 
HD 217089           &    --7.441 $\pm$ 0.150 &    --7.684  $\pm$ 0.001 & \phantom{$\bullet$} \phantom{$\circ$} \phantom{$\star$} \\ 
HD   9637           &    --1.879 $\pm$ 0.169 &    --2.361  $\pm$ 0.002 & \phantom{$\bullet$} \phantom{$\circ$} \phantom{$\star$} \\ 
HD  21269           &   --15.676 $\pm$ 0.262 &   --12.602  $\pm$ 0.002 & \phantom{$\bullet$} $\circ$            \phantom{$\star$} \\ 
HD  19267           &      3.387 $\pm$ 0.153 &      2.697  $\pm$ 0.001 & \phantom{$\bullet$} \phantom{$\circ$} $\star$ \\ 
HD  13882           &   --28.545 $\pm$ 0.170 &   --29.038  $\pm$ 0.001 & \phantom{$\bullet$} \phantom{$\circ$} \phantom{$\star$} \\ 
HD 189879           &   --10.391 $\pm$ 4.430 &   --29.532  $\pm$ 0.001 & $\bullet$           \phantom{$\circ$} $\star$ \\ 
HD 195375 $\ast$    &    --9.352 $\pm$ 0.154 &   --10.276  $\pm$ 0.001 & \phantom{$\bullet$} $\circ$           \phantom{$\star$}\\ 
HD 221232           &   --29.894 $\pm$ 0.158 &   --30.649  $\pm$ 0.001 & \phantom{$\bullet$} \phantom{$\circ$} \phantom{$\star$}\\
HD 219925           &   --14.596 $\pm$ 2.837 &   --22.916  $\pm$ 0.001 & $\bullet$           \phantom{$\circ$} $\star$ \\
HD    278 $\ast$    &   --26.143 $\pm$ 7.736 &   --60.526  $\pm$ 0.003 & $\bullet$            \phantom{$\circ$} \phantom{$\star$} \\  
HD  11519           &   --10.625 $\pm$ 0.177 &   --11.260  $\pm$ 0.001 & \phantom{$\bullet$} \phantom{$\circ$} \phantom{$\star$} \\  
TYC 2813-1979-1     &   --16.201 $\pm$ 0.338 &   --17.170  $\pm$ 0.002 & \phantom{$\bullet$} \phantom{$\circ$} \phantom{$\star$} \\  
BD+42  3220         &   --18.523 $\pm$ 0.219 &   --19.298  $\pm$ 0.053 & \phantom{$\bullet$} \phantom{$\circ$} \phantom{$\star$} \\  
BD+44  3114         &   --17.302 $\pm$ 0.170 &   --15.216  $\pm$ 0.055 & \phantom{$\bullet$} $\circ$           \phantom{$\star$} \\  
TYC 3136-878-1      &      0.922 $\pm$ 0.985 &      1.034  $\pm$ 0.053 & \phantom{$\bullet$} \phantom{$\circ$} \phantom{$\star$} \\  
HD  40509           &    --1.223 $\pm$ 0.310 &    --1.656  $\pm$ 0.005 & \phantom{$\bullet$} \phantom{$\circ$} \phantom{$\star$} \\  
HD  41710           &    --6.812 $\pm$ 0.269 &    --7.917  $\pm$ 0.002 & \phantom{$\bullet$} \phantom{$\circ$} \phantom{$\star$} \\  
HD  40655           &                        &      7.332  $\pm$ 0.001 & \phantom{$\bullet$} \phantom{$\circ$} \phantom{$\star$} \\  
HD  45879           &      5.902 $\pm$ 0.224 &      7.047  $\pm$ 0.001 & \phantom{$\bullet$} $\circ$           $\star$ \\ 
HD  55077           &   --25.729 $\pm$ 0.514 &   --25.499  $\pm$ 0.003 & \phantom{$\bullet$} \phantom{$\circ$} $\star$ \\ 
HD  61107           &     12.126 $\pm$ 0.738 &     10.476  $\pm$ 0.001 & \phantom{$\bullet$} \phantom{$\circ$} $\star$ \\ 
HD  63856           &     22.402 $\pm$ 0.206 &     19.755  $\pm$ 0.001 & \phantom{$\bullet$} $\circ$           $\star$ \\ 
\hline 
\end{tabular}
}
\tablefoot{Stars with $\text{an asterisk}$ are in the Washington Double Star Catalog. $\text{Filled circles}$  indicate stars with Gaia DR2 radial velocity errors $>$ 1 \kms. $\text{Open circles}$  represent stars with Gaia DR2 radial velocities that differ by more than 5$\sigma$ from our observed \vrad. $\text{The five-pointed star}$  indicates stars that were identified as possible binary stars from proper motion anomalies by \cite{Kervella}.}
\end{table}

Stars HD 195375 and HD 278 are listed in the Washington Double Star Catalog \citep{WashingtonCat}. We checked the Gaia EDR3 astrometric parameters of these two binary systems. We found that HD 278 and its companion have different parallaxes, so that these stars are probably in a visual double
system but not in a physical one, as suggested by \citet{Muller}. 
In contrast, HD 195375 and its companion have consistent parallaxes, which implies that 
they are in a physical binary system.
As binarity is a possible cause of radial velocity variability, 
we compared our measured radial velocities with those
of Gaia DR2 \citep{GaiaDR2}. 
The brightness of our stars implies that the error on the radial velocity due to photon noise is smaller than 1\,\kms\ \citep{Sartoretti}.
As shown in Table \ref{tab:gaiarv}, six stars 
have a larger error, which means that they are very likely radial
velocity variables. 
The Gaia radial velocity of eight stars differs
by more than 5$\sigma$ from our measured velocity. They are again very likely radial velocity variables. 
Ten stars have been identified as probable binary stars by \cite{Kervella} from the comparison between \textsc{Hipparcos} \citep{Hipparcos} and Gaia DR2 proper motions. 
We note that stars HD 192045 and HD 213036 show all the properties mentioned above. This strongly suggests that they are binary stars.  
It is likely that the end-of-mission Gaia data, which will combine astrometry,
epoch photometry, and radial velocities,
will allow us to determine the orbits of the confirmed binaries.

\section{Kinematics and Galactic orbits}

We characterised the stellar orbital parameters with the Galpot code\footnote{\href{https://github.com/PaulMcMillan-Astro/GalPot}{https://github.com/PaulMcMillan-Astro/GalPot}} \citep{McMillan,DehnenBinney}
using the stellar coordinates, radial velocities, Gaia DR2 distances, and proper motions. 
We derived the stellar coordinates and velocity components in the galactocentric cylindrical (R, z, $\varphi$, v$_R$, v$_Z$, v$_\varphi$) and Cartesian systems (X, Y, Z, v$_X$, v$_Y$, v$_Z$). We derived the minimum and maximum cylindrical (R$_{min}$, R$_{max}$) and spherical (r$_{min}$, r$_{max}$) radii, the eccentricity (e=(r$_{max}$ -- r$_{min}$)/(r$_{max}$ + r$_{min}$)), the maximum height above the Galactic plane (Z$_{max}$), the total energy (E), and the z-component of the angular momentum (L$_Z$).
We found that all stars have typical disc kinematics. All stars have prograde motions, and the eccentricity (e $<$ 0.11) for 25 out of 26 stars (e$\sim$0.2 for HD 278) is very low. The maximum height above the Galactic plane is lower than 400 pc for 25 out of 26 stars (Z$_{max}\sim 600$ pc for TYC 2813-1979-1). 

\section{Discussion}

\subsection{Photometric logg versus spectroscopic logg}\label{logg}

In order to verify the values of \logg\ that we obtained from Gaia photometry and parallaxes, we derived surface gravities by imposing the ionisation equilibrium of \ion{the Fe}{I} and \ion{Fe}{II} lines. As shown in 
Table \ref{loggspec}, photometric and spectroscopic \logg\ are compatible within 0.1 dex for ten stars, while the difference between photometric and spectroscopic gravities is $\mathrm{-0.6\leq \Delta\logg \leq 0.9}$ dex for the remaining stars.
We deduced the corresponding stellar masses from spectroscopic gravities by inverting Eq. \ref{eqlogg}. We obtained that the masses should be $\mathrm{0.7\leq M/M_{\odot}\leq18.9;}$  two stars have $\mathrm{M<0.8\ M_\odot}$. These very low masses are incompatible with stellar evolutionary models (unless we assume that they are older than 14 Gyr). We therefore conclude that the spectroscopic gravities are not reliable for the majority of these stars. The reason for the observed discrepancy is not trivial because for solar metallicity stars, we expect spectroscopic and photometric approaches to be equivalent (see the discussion in 
\citealt{Mucciarelli&Bonifacio}). As the difference in \logg\ reflects the difference between \ion{Fe}{I} and \ion{Fe}{II} abundances, the discrepancy for the most highly
rotating stars could be due to a bias in the line selection. Stellar rotation allowed us to detect only the strongest and consequently most saturated \ion{Fe}{II} lines. However, the same discrepancy is found for some stars with low rotational velocity, therefore some other effect must be responsible for it. The observed scatter in $\Delta$\logg\ for stars with \vseni\,$<$\,10\ \kms\ seems to suggest that this effect is due to inadequacies of the adopted physics, in particular, the assumption of 1D geometry and LTE, as discussed for metal-poor giants in \citet{Mucciarelli&Bonifacio}. 
We confirmed that non-LTE (NLTE) effects under the assumption of 1D geometry are not sufficient to solve the \logg\ discrepancy. We applied the NLTE corrections provided by \citet{Bergemann}\footnote{\href{http://nlte.mpia.de/}{http://nlte.mpia.de/}} to the \ion{Fe}{I} lines of the star HD 191066, for which we obtained [FeI/FeII]=0.26 dex using photometric \logg. We found that the mean correction value was 0.005 dex with a maximum value of 0.06 dex for \ion{the Fe}{I} line at 5956.693 \AA.
This means that applying this NLTE correction to \ion{Fe}{I} the imbalance would be even worse.

\begin{table}[t]
\caption{Comparison between photometric and spectroscopic \logg\ and derived stellar masses.} 
\label{loggspec}
\centering
\resizebox{\hsize}{!}{
\begin{tabular}{l r r r r r}
\hline\hline
Star & \logg$_{phot}$ & \logg$_{spec}$ & $\Delta$\logg & Mass & Mass$_{spec}$ \\
 & dex & dex & dex & \Msol & \Msol \\     
\hline
HD 192045       &  2.91 & 3.21 & -0.30 & 3.0 & 6.1   \\
HD 191066       &  3.01 & 3.51 & -0.50 & 3.0 & 9.5   \\
HD 205732       &  2.54 & 3.14 & -0.60 & 4.0 & 16.1  \\
HD 213036       &  1.94 & 2.44 & -0.50 & 6.0 & 18.9  \\
HD 217089       &  2.61 & 2.61 &  0.00 & 3.5 & 3.5   \\
HD   9637       &  2.83 & 2.73 &  0.10 & 3.5 & 2.8   \\
HD  21269       &  1.83 & 1.73 &  0.10 & 6.0 & 4.7   \\
HD  19267       &  2.50 & 1.90 &  0.60 & 4.0 & 1.0   \\
HD  13882       &  2.45 & 1.85 &  0.60 & 4.0 & 1.0   \\
HD 189879       &  2.08 & 2.08 &  0.00 & 6.0 & 6.0   \\
HD 195375       &  2.56 & 1.86 &  0.70 & 3.5 & 0.7   \\
HD 221232       &  2.46 & 1.96 &  0.50 & 4.0 & 1.3   \\
HD 219925       &  2.61 & 3.11 & -0.50 & 3.5 & 11.1  \\
HD    278       &  2.59 & 2.59 &  0.00 & 3.5 & 3.5   \\
HD  11519       &  2.35 & 2.35 &  0.00 & 4.0 & 3.9   \\
TYC 2813-1979-1 &  2.67 & 2.67 &  0.00 & 3.5 & 3.5   \\
BD+42  3220     &  3.02 & 3.02 &  0.00 & 2.5 & 2.5   \\
BD+44  3114     &  2.77 & 2.77 &  0.00 & 3.0 & 3.0   \\
TYC 3136-878-1  &  2.96 & 2.96 &  0.00 & 3.0 & 3.0   \\
HD  40509       &  2.62 & 2.22 &  0.40 & 3.5 & 1.4   \\
HD  41710       &  2.03 & 1.63 &  0.40 & 6.0 & 2.4   \\
HD  40655       &  2.21 & 1.71 &  0.50 & 5.0 & 1.6   \\
HD  45879       &  1.90 & 1.00 &  0.90 & 6.0 & 0.7   \\
HD  55077       &  2.59 & 2.29 &  0.30 & 3.5 & 1.7   \\
HD  61107       &  2.56 & 2.36 &  0.20 & 4.0 & 2.5   \\
HD  63856       &  2.67 & 3.17 & -0.50 & 3.5 & 11.1  \\
\hline
\end{tabular}
}
\end{table}

\subsection{Chemical composition}

\begin{figure}
    \centering
    \includegraphics[width=\hsize]{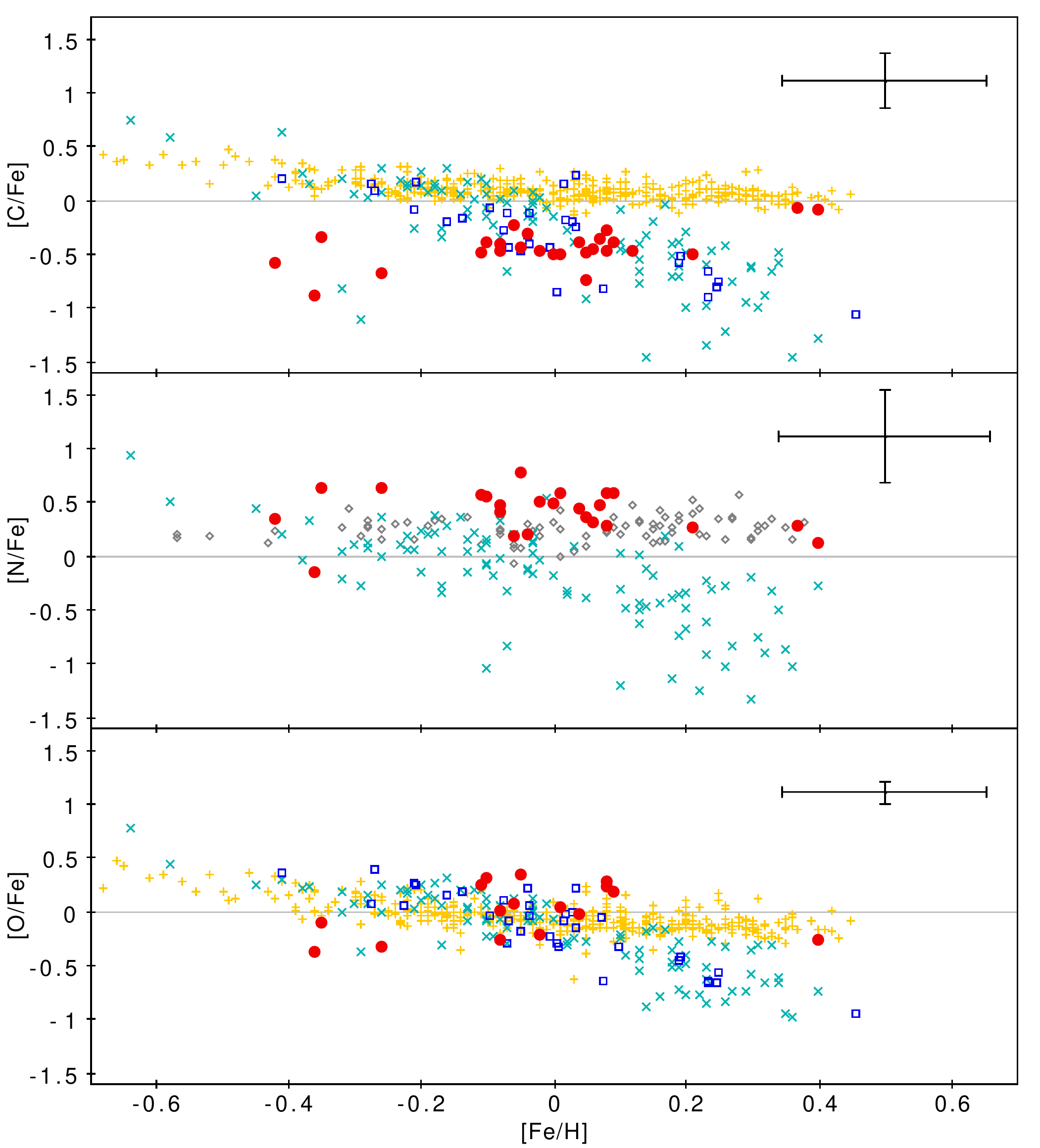}
    \caption{[C/Fe], [N/Fe] and [O/Fe] abundances as a function of [Fe/H]. Comparison with targets from \cite{Takeda} (cyan crosses), \cite{Royer} (blue squares), \cite{DelgadoMena} (yellow crosses) and \cite{Ecuvillon} (grey diamonds).}
    \label{cno_fe}    
\end{figure}

\begin{figure}
    \centering
    \includegraphics[width=\hsize]{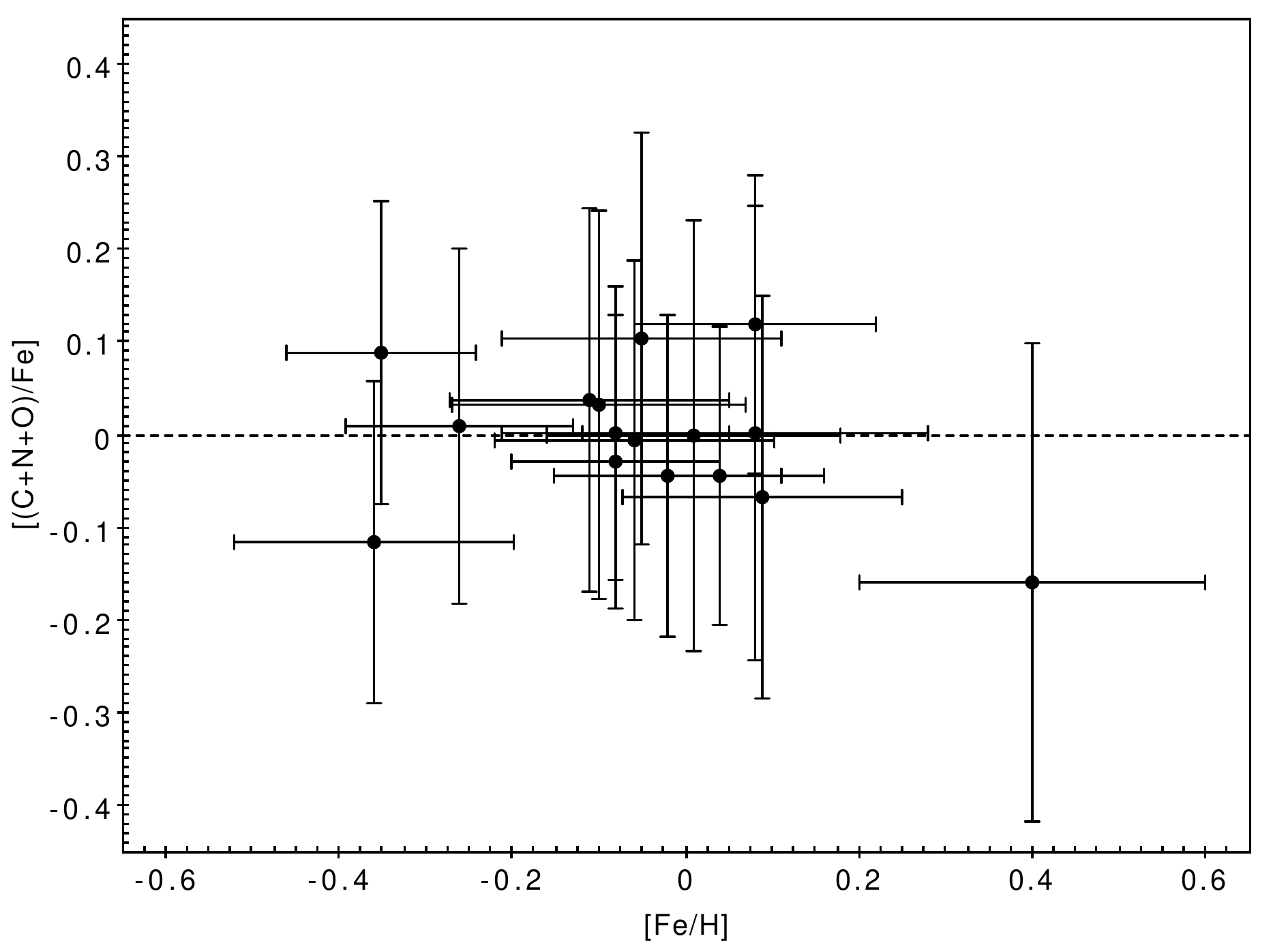}
    \caption{[(C+N+O)/Fe] abundance ratios as a function of [Fe/H]. }
    \label{cpnpo_fe}    
\end{figure}

\subsubsection{C,N,O}\label{cno}

In Fig. \ref{cno_fe} we show our  measured C, N and O abundance ratios. 
The [C/Fe] ratios are lower than solar for all stars in our sample (<[C/Fe]>=--0.44 dex and $\sigma$=0.17), with the exception of two stars with [Fe/H]$\sim$0.4 dex that show a  [C/Fe] of about zero.
Our interpretation is that all the stars with sub-solar [C/Fe] show material in the photosphere
that has been mixed with material that has experienced nuclear hydrogen
burning through the CNO cycle. This interpretation is supported by the
super-solar [N/Fe] ratios. The 
[(C+N+O)/Fe] ratio is shown in Fig. \ref{cpnpo_fe}. This quantity is nearly constant and close to the solar value within the error bars. This indicates that the underabundances in C and the overabundances in N result from pure H-burning via the CNO cycle, and that no He-burning products have yet been transported to the stellar surface.
The value of [C/Fe] of a given star is probably independent of
the original [C/Fe] value that characterised the star on the main sequence, 
but depends only on the amount of mixing.
This is supported by the fact that there is no clear trend of 
[C/Fe] with metallicity.
If our interpretation is correct, the two more metal-rich stars
of our sample are mixed very little with respect to the others.
Some mixing is only suggested by a slight enhancement of [N/Fe]
in both stars. These two stars belong to the stars with the highest
gravity in the sample, which is in line with the notion
of little or no mixing. We caution, however, that there
are stars with similarly high log g that have a low [C/Fe].
There clearly is no one-to-one correspondence between 
surface gravity and mixing. 

It is interesting to compare our results with those in the literature. 
We selected two samples of CNO abundances in  A-type stars ( 
 \citet{Takeda} and \citet{Royer}). The Takeda and Royer samples both consists of A-type main-sequence stars. Royer stars are also characterised by \vseni\,$\leq$\,65 \kms. 
We also added samples of  CNO samples in FGK dwarf stars with a
metallicity comparable to that of our sample: 
\citet{DelgadoMena} provided C and O abundances for  370 FGK dwarfs stars,
and \citet{Ecuvillon} provided N abundances for 91 solar-type stars.
The top panel of Fig.\,\ref{cno_fe} clearly shows that
the majority of our stars have lower [C/Fe] than the FGK dwarfs (yellow crosses), 
with the exception of the two stars with the highest metallicity, which appear to be
quite compatible. 
It is striking that the [C/Fe] abundance in A-type stars displays
a decreasing trend with increasing metallicity for the samples
of \citet{Takeda} (cyan crosses) and \citet{Royer} (blue squares).
This trend is at odds with the flat trend shown by FGK stars.
This trend in A stars has been noted and discussed by \citet[][see also references therein]{Takeda}. A stars show the phenomenon of chemical peculiarities (CP stars), which gives rise to many
sub-classes of CP stars \citep[see][and references therein]{cpstars}. 
The most popular explanation of the chemical peculiarities for
most classes of CP stars is diffusion \citep[see e.g.][and references therein]{richer},
possibly in the presence of rotational mixing \citep{talon}.
As pointed out by \citet{Takeda}, this anti-correlation of [C/Fe]
with [Fe/H] can be understood if the mechanism causing the chemical
peculiarities acts in opposite directions for CNO and Fe.
We do not detect any chemical peculiarities, except for the 
CNO pattern expected from mixing on the RGB and Ba (see Sect.\,\ref{ncapture}). This
strongly suggests that these peculiarities, even if they were present when
the star was on the main sequence, are erased as the star evolves 
to the RGB by the onset of convective mixing  as the star
cools and its atmosphere is no longer in radiative equilibrium,  
as was the case while the star was on the main sequence.

\begin{figure}
    \centering
    \includegraphics[width=\hsize]{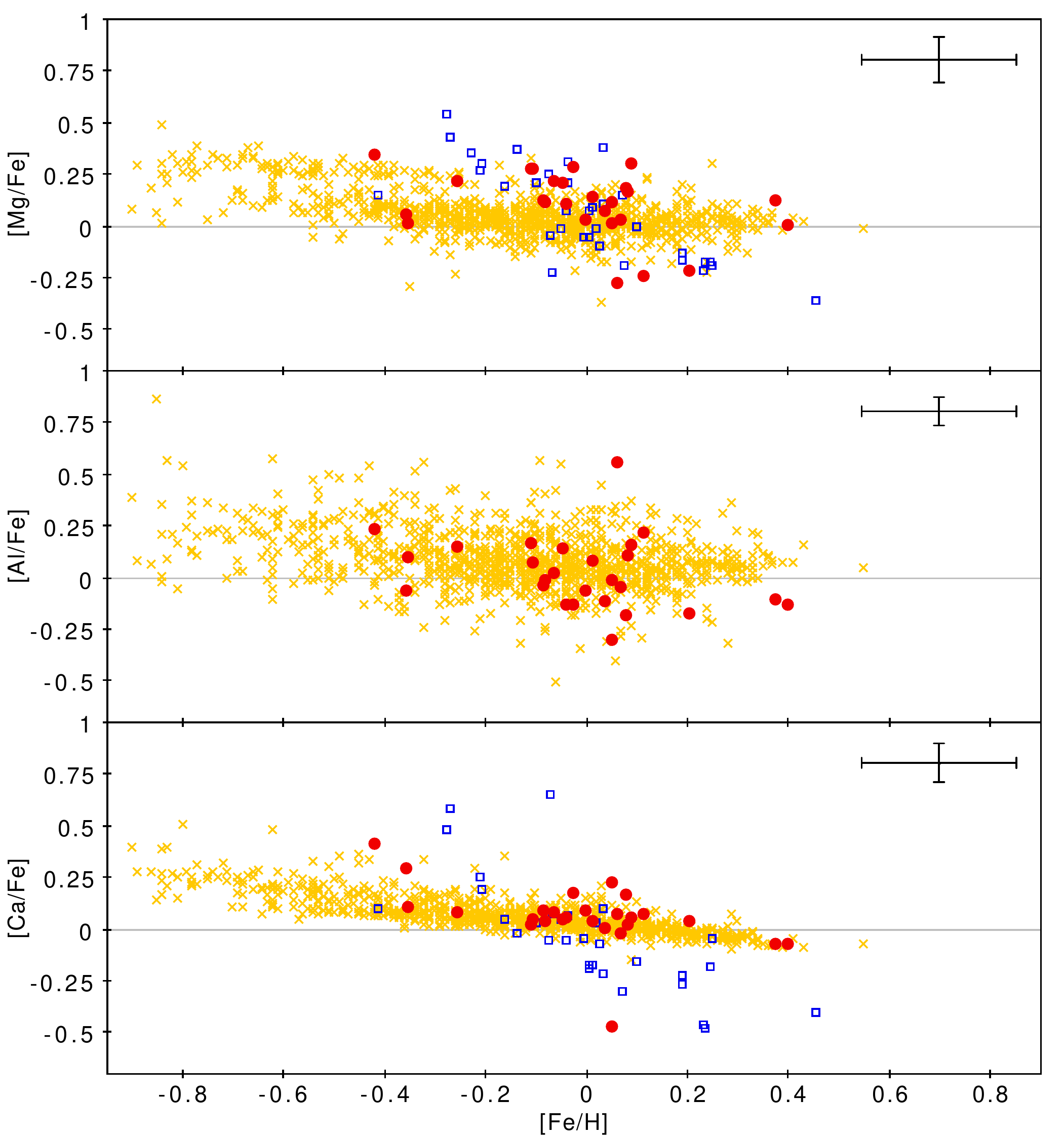}
    \caption{[Mg/Fe], [Al/Fe], and [Ca/Fe] abundances as a function of [Fe/H]. Comparison with targets from \cite{Adibekyan} (yellow crosses) and \cite{Royer} (blue squares).
    }
    \label{mgalca_fe}    
\end{figure}

\subsubsection{Mg, Al, and Ca}

Figure \ref{mgalca_fe} shows Mg, Al, and Ca abundance ratios as a function of [Fe/H]. Our results are compared to the analysis by \cite{Adibekyan} (yellow crosses) and \cite{Royer} (blue squares). 
The Adibekyan sample consists of F, G, and K dwarf stars, which means that they have lower masses than our sample stars, but \teff\ is similar. 
Our derived [Mg/Fe], [Al/Fe], and [Ca/Fe] abundance ratios appear to be in line with the results obtained by other authors. 
The dispersion in [Mg/Fe] (<[Mg/Fe]>=0.10 dex and $\sigma$=0.16), [Al/Fe] (<[Al/Fe]>=0.02 dex and $\sigma$=0.17) and [Ca/Fe] (<[Ca/Fe]>=0.06 dex and $\sigma$=0.15) is larger than is expected from our estimated errors. In the case of [Mg/Fe], we note that the \citeauthor{Adibekyan} and \citeauthor{Royer} stars also show a large dispersion despite the high quality of spectra. The observed scatter is therefore probably intrinsic. 
The [Mg/Fe] ratio of two stars in our sample appears to be compatible with the values of stars in the upper sequence in the \citeauthor{Adibekyan} sample. They were  labelled thick-disc stars by the authors. However, the kinematics of these two stars clearly shows that they are thin-disc stars. 
A purely chemical selection is not sufficient to distinguish between thin- and thick-disc stars,
as has been pointed out by several authors \citep[see e.g.][]{Franchini2020,Romano}. 
Specifically, \citet{Romano} reported that  only 25\% of the high-$\alpha$
stars in their sample can be classified kinematically as belonging to the thick-disc population. It is therefore not surprising
that we find thin-disc high-$\alpha$ stars.
Some unexpected results were found for stars HD 278 and HD 21269. 
Star HD 278 shows a higher Al abundance than the other stars in the sample ([Al/Fe]=0.55 dex, $\sigma$=0.24). 
Star HD 21269 instead shows a higher than solar Mg abundance ([Mg/Fe]=0.11 dex, $\sigma$=0.36) and lower than solar Al and Ca abundance ([Al/Fe]=--0.30 dex, $\sigma$=0.11; [Ca/Fe]=--0.47, $\sigma$=0.26). 
These stars rotate rapidly (\vseni$\sim$22 \kms\ for HD 278 and \vseni$\sim$16 \kms\ for HD 21269)  and the uncertainty on the abundances is large. As discussed in Sect. \ref{logg}, rotation can affect the estimate of the EW of the lines, which may lead to an incorrect abundance estimate. However, we note that some stars in the \citeauthor{Adibekyan} sample show the same Al abundance. This implies that it is possible for a star to have such  a low [Al/Fe] ratio. 

\begin{figure}
    \centering
    \includegraphics[width=\hsize]{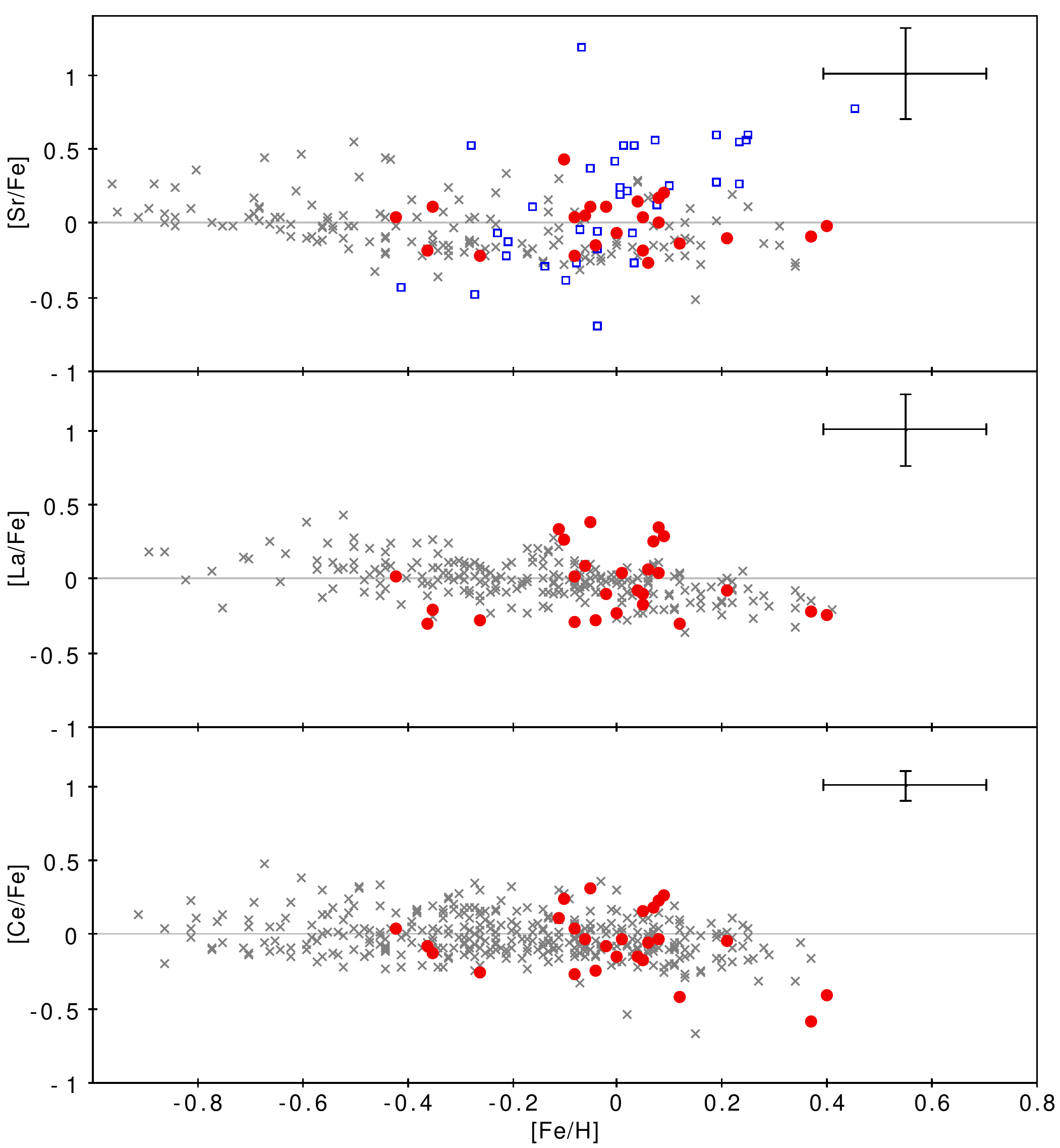}
    \caption{[Sr/Fe], [La/Fe], and [Ce/Fe] abundances as a function of [Fe/H]. Comparison with targets from \cite{Battistini} (grey crosses) and \cite{Royer} (blue squares).}
    \label{srlace_fe}    
\end{figure}

\begin{figure}
    \centering
    \includegraphics[width=\hsize]{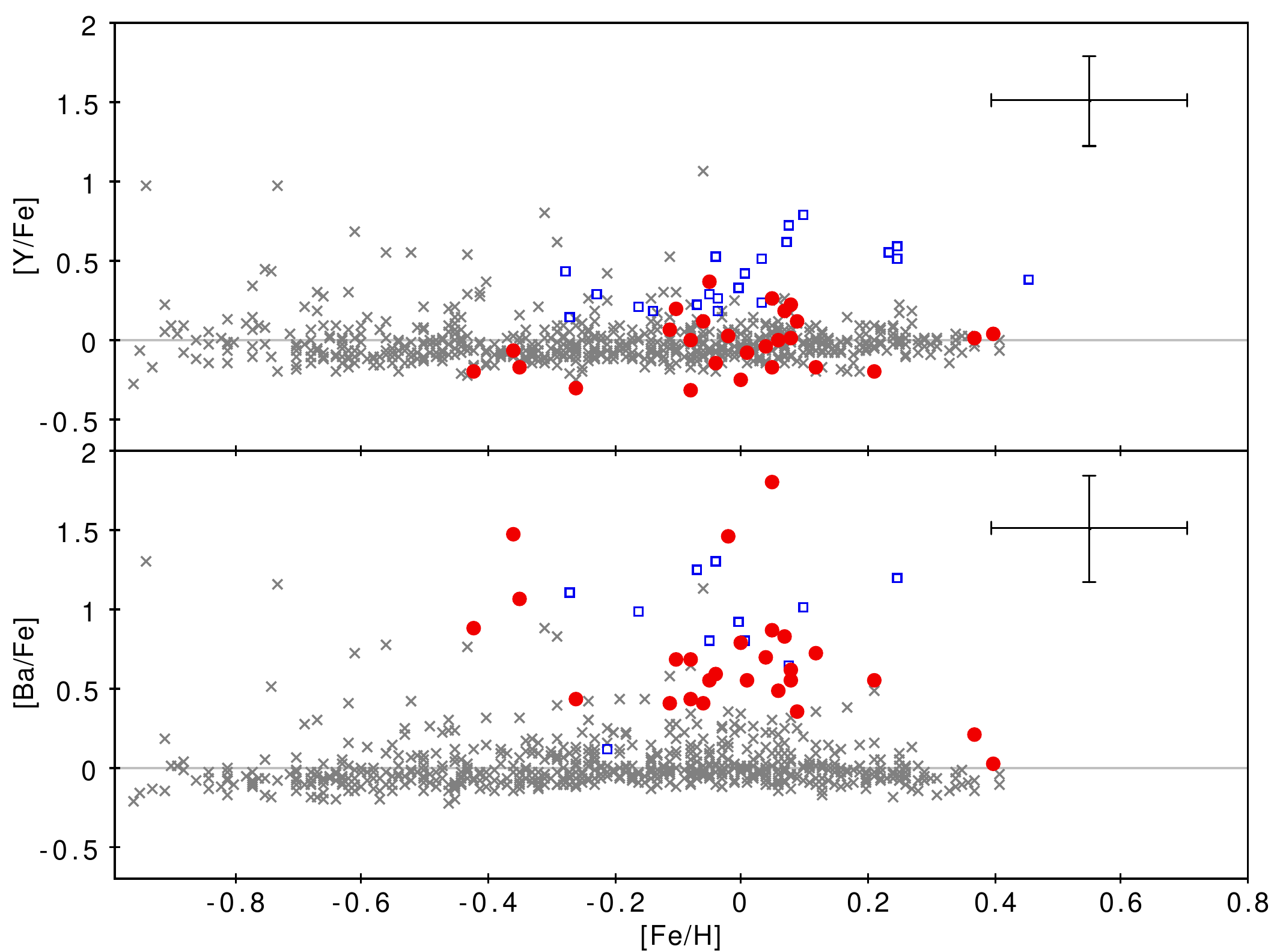}
    \caption{[Y/Fe] and [Ba/Fe] abundances as a function of [Fe/H]. Comparison with targets from \cite{Bensby} (grey crosses) and \cite{Royer} (blue squares).}
    \label{yba_fe}    
\end{figure}

\begin{figure}
    \centering
    \includegraphics[width=\hsize]{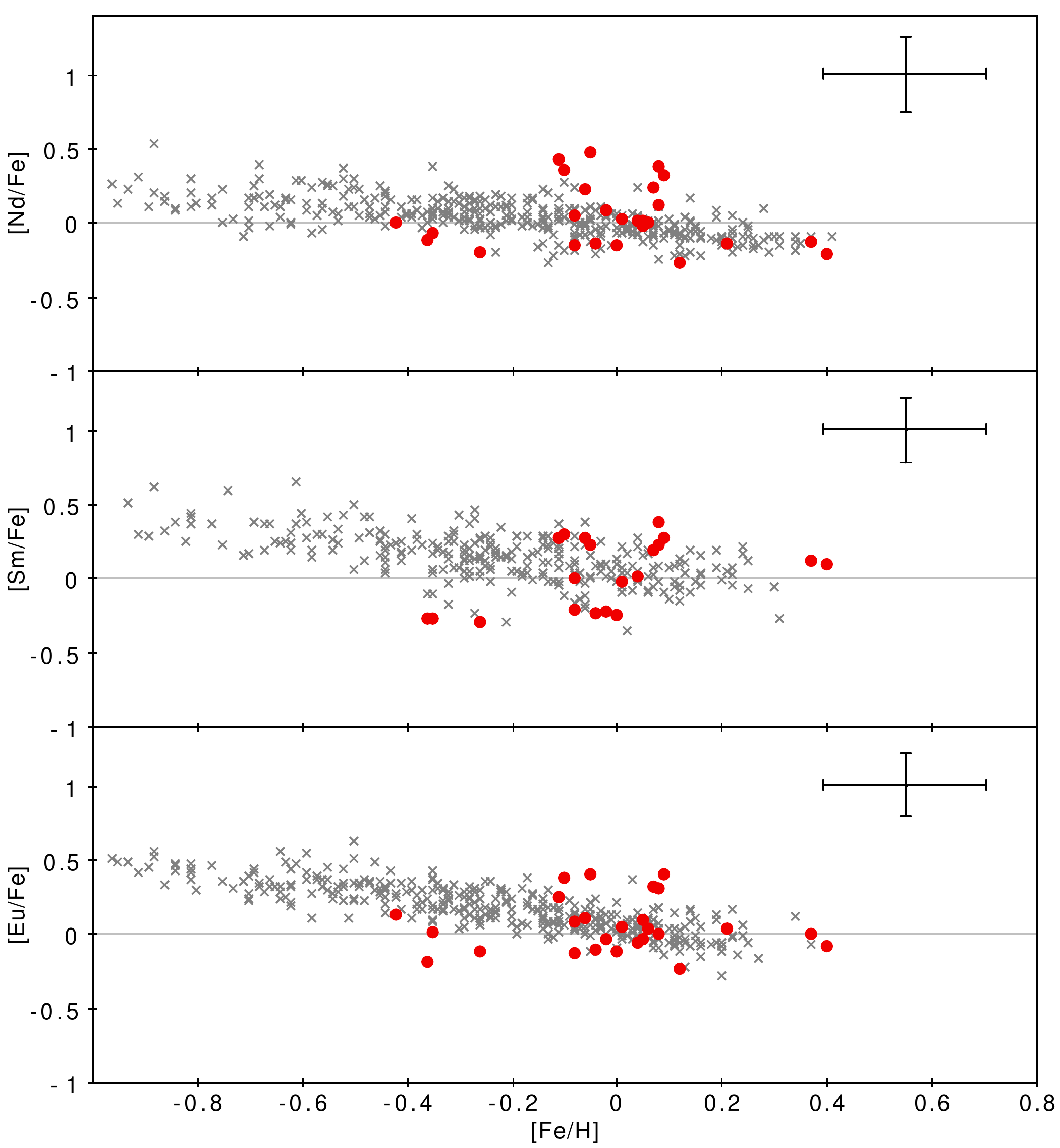}
    \caption{[Nd/Fe], [Sm/Fe], and [Eu/Fe] abundances as a function of [Fe/H]. Comparison with targets from \cite{Battistini} (grey crosses).}
    \label{ndsmeu_fe}    
\end{figure}

\begin{figure}
\centering
\includegraphics[width=\hsize]{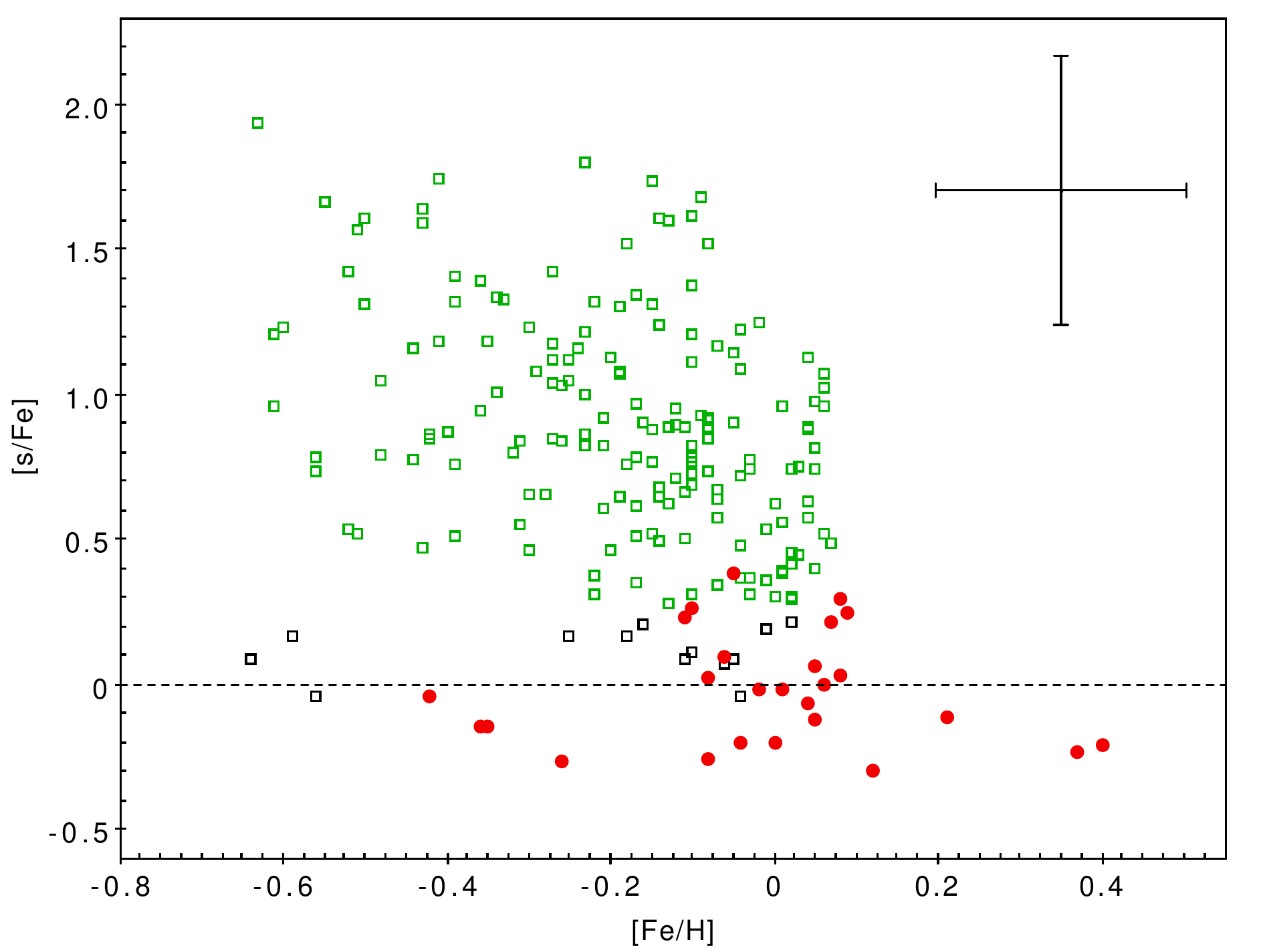}
\caption{Mean abundance ratio [s/Fe] of s-process elements ([Y/Fe], [La/Fe], [Ce/Fe], and [Nd/Fe]) as a function of [Fe/H]. Red circles represent the [s/Fe] ratio for the stars analysed in this work. Green squares represent barium stars analysed in \cite{deCastro}. Black squares represent the targets rejected as barium stars in \cite{deCastro}.}
\label{sfe_feh}
\end{figure}

\subsubsection{Neutron-capture elements \label{ncapture}}

In Figs. \ref{srlace_fe}, \ref{yba_fe}, and \ref{ndsmeu_fe} we show the abundances of several neutron-capture elements. The 
Sr abundance ratio is compared to the analysis by \citet{Royer} (blue squares) and \cite{Battistini} (grey crosses), and the [Y/Fe] and [Ba/Fe] ratios are compared to the results of the analysis by \citet{Royer} (blue squares) and \citet{Bensby} (grey crosses). 
The La, Ce, Nd, Sm, and Eu abundances ratios are compared to the analysis by \citet{Battistini} (grey crosses). The \citet{Bensby} sample consists of F and G dwarf and subgiant stars, and
 \citet{Battistini} provided abundances of several n-capture elements for the same stars. 
We observe that in the case of n-capture elements, the abundances ratios we measured are also in line with the results found by other authors. A remarkable result is the 
Ba abundance, which is higher than solar for the majority of stars in our sample. The same result has been found for main-sequence stars by \citet{Royer}. The NLTE correction for Ba lines provided by \cite{Korotin} would decrease the Ba abundances by 0.1 dex, which is insufficient to match the observed values with other n-capture abundances. We exclude the possibility that an atmospheric phenomenon could explain the observed Ba abundance: if the Ba overabundance were due to atmospheric processes during the main-sequence phase, it would be erased by mixing when the star evolves to the red giant phase.

In Fig. \ref{sfe_feh} we compare our results to the s-process abundance ratios [s/Fe], that is, the mean abundance ratio [X/Fe] of s-process elements [Y/Fe], [La/Fe], [Ce/Fe], and [Nd/Fe] of barium stars derived by \cite{deCastro}.
We observe that three stars in our sample have [s/Fe] above 0.25 dex, which is the lower limit for [s/Fe] in Ba stars according to \cite{deCastro}, and only one star (HD 63856) has [s/Fe]=0.38 dex. These low values of the [s/Fe] ratio are compatible with those of mild Ba stars, which have weaker s-process enhancements than classical barium stars, only few or no anomalous molecular band strengths \citep{Eggen,MorganKeenan}, and no carbon enrichment \citep{Sneden}. 
However, we stress that the \ion{Ba}{II} abundance is strongly dependent on microturbulent velocity. We note that an increase in microturbulence of 0.6 \kms\ decreases the Ba abundance by 0.6 dex for star HD 55077. This effect was first observed by \cite{HylandMould}, who showed that high microturbulent velocities might cause the \ion{Ba}{II} resonance lines to become anomalously strong in stars with solar s-process elements. In our case, we have no
observation that supports such a high microturbulence, which would also
affect the abundances of other elements.
There probably is no strong evidence 
implying that stars in our sample are mild Ba stars, but the high [Ba/Fe] ratios
are puzzling and unexplained.

In their study of dwarf stars in five open clusters and one star-forming region, \citet{Baratella} found an enhancement in Ba similar to what we have found in our stars. Remarkably, Ba is more enhanced for the younger clusters. The Ba enhancement is accompanied by mild Y enhancement of the order of 0.2 dex. A larger enhancement is again found for the younger clusters. These two results are consistent with our findings. At face value, our derived [Y/Fe] abundance ratios are $\sim$ 0.2 dex, but we do not give much weight to the Y enhancement as it is consistent with [Y/Fe]=0 within the errors.  
\citet{Baratella} examined several causes for this Ba anomaly, including the role of magnetic fields. They failed to propose any convincing explanation of their observations, however.  
We therefore conclude that the increase in Ba that we observe may be related to what has been observed in other young stars.

\subsubsection{Chemical abundances and rotational velocities}

We investigated the presence of  trends between elemental abundances and rotational velocity in our sample.
To limit evolutionary effects, we considered only stars with metallicity in the range $\mathrm{-0.1<[Fe/H]<0.1}$ .
We found that for only four elements ([O/H], [Ca/H], [Ba/H], and [Eu/H]) 
does non-parametric Kendall's $\tau$ test provides a correlation probability higher than 95\% with \vseni. 
However, parametric fitting did not confirm any 
trend between these quantities.
We therefore conclude that the correlations are not significant. 

\begin{figure}
    \centering
    \includegraphics[width=\hsize]{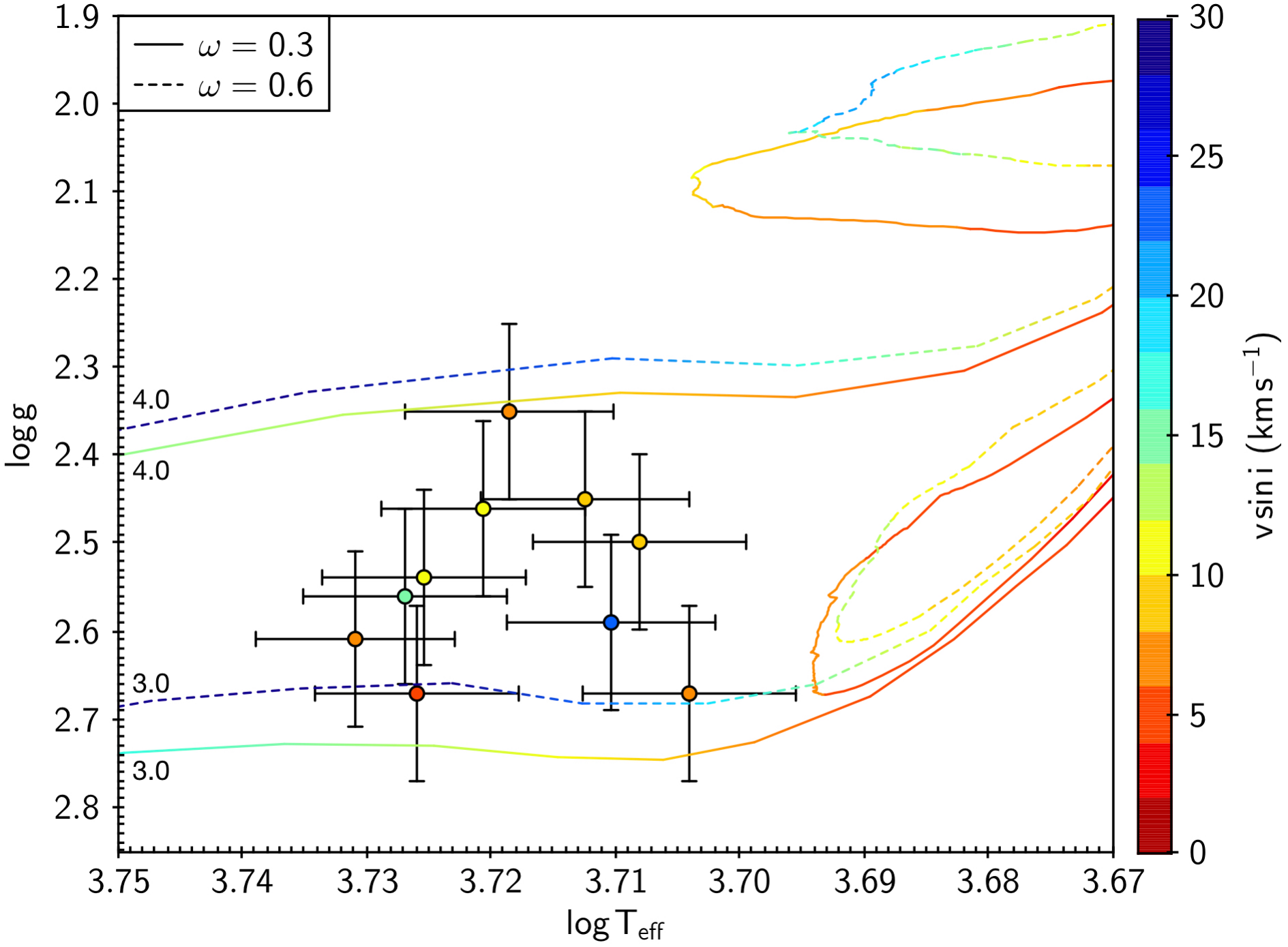}
    \includegraphics[width=\hsize]{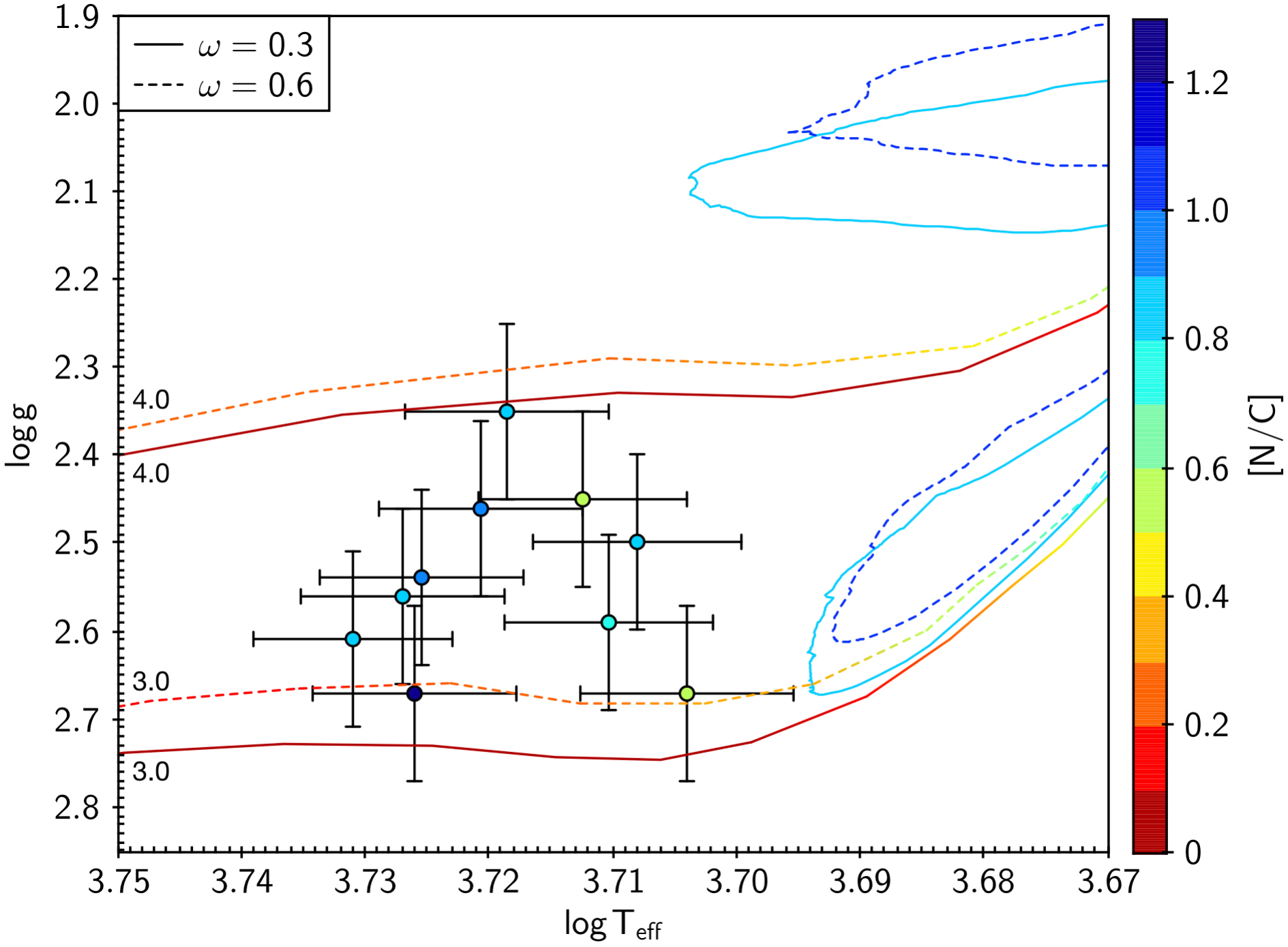}
    \caption{Log\,\teff\  vs log\,g diagram for the \cite{Georgy} evolutionary tracks with Z = 0.014, M = 3 \Msol, 4 \Msol, and $\omega$ = 0.3 (solid lines) and 0.6 (dashed lines). Stars in the sample with $\mathrm{-0.1<[Fe/H]<0.1}$ and M = 3.5 \Msol, 4 \Msol\ are shown with error bars. Upper panel: Colour index indicates the equatorial velocity for evolutionary tracks and the observed \vseni\ for sample stars. Lower panel: Colour index indicates the [N/C] abundance ratio.}
    \label{logteff_logg}    
\end{figure}

\begin{figure}
    \centering
    \includegraphics[width=\hsize]{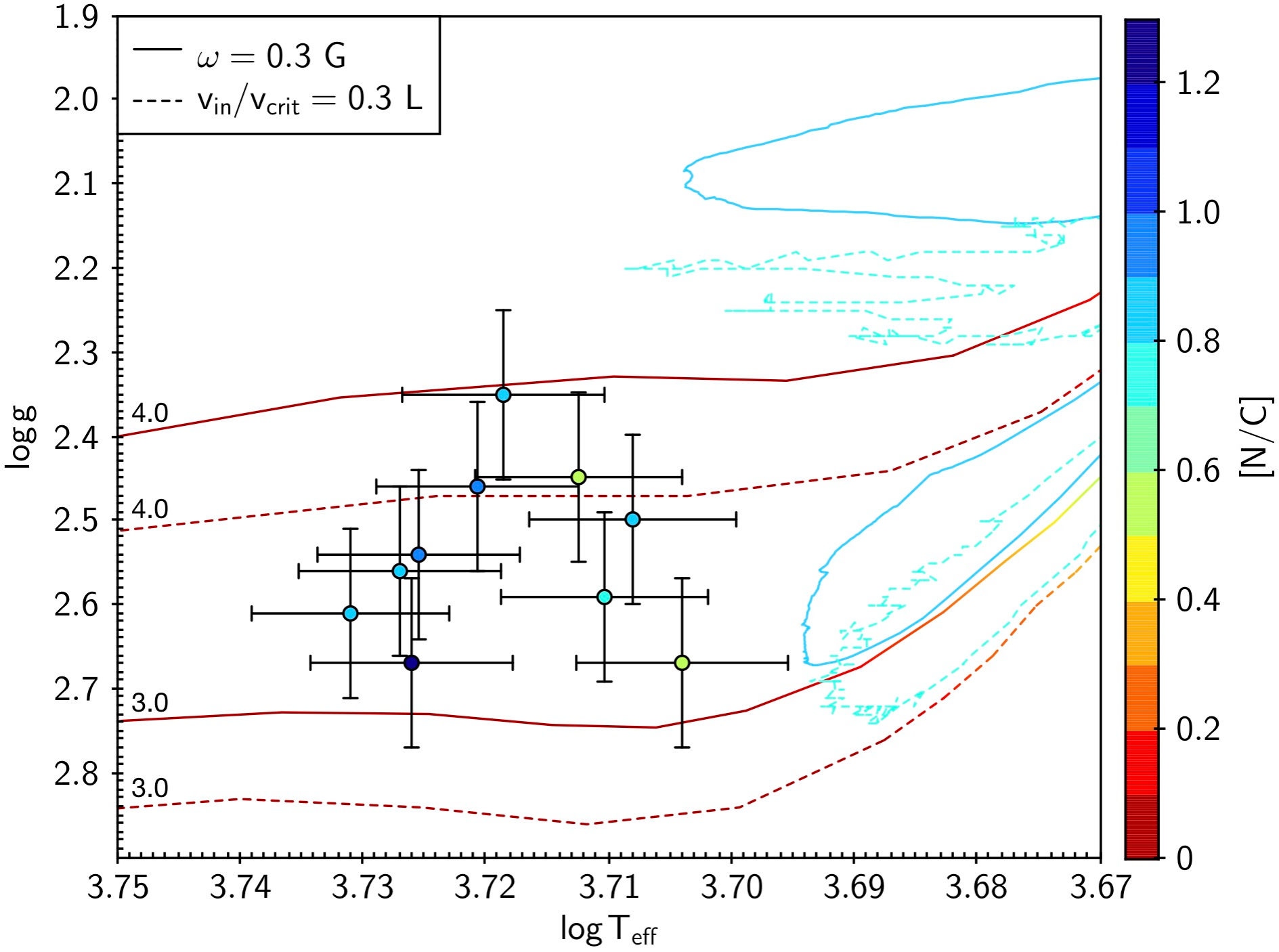}
    \caption{Comparison between the \cite{Georgy} (solid lines) and \cite{Lagarde} (dashed lines) evolutionary tracks with rotation at $\mathrm{Z = 0.014}$ for M = 3 \Msol, 4 \Msol. Stars in the sample with M = 3.5 \Msol, 4 \Msol\ and $\mathrm{-0.1<[Fe/H]<0.1}$ are shown with error bars. The colour index indicates the [N/C] abundance ratios.}
    \label{logteff_logg_g_l}    
\end{figure}

\subsection{Comparison with models}

We compared our results with the predictions of two sets of stellar evolution models including the effects of rotation: the models of \cite{Georgy}, which were computed with the Geneva stellar evolution code (GENEC), and the models of \cite{Lagarde}, which were computed with the code STAREVOL \citep{MowlaviForestini,Siess,Palacios03,Palacios06,Decressin}; see \cite{Lagarde} for a comparison between these two sets of models.

Because the observed rotational velocities depend on the inclination angle of the stars along the line of sight, a direct comparison between observed and predicted rotational velocities is not possible. However, the observed values of \vseni\ are lower limits for the actual surface rotation of stars. As shown in the upper panel of Fig. \ref{logteff_logg}, the \vseni\ we obtained for stars with 3.5 \Msol\ and 4 \Msol\ in the metallicity range $\mathrm{-0.1<[Fe/H]<0.1}$ ($5<\vseni<22$ \kms) are compatible with the rotational velocity values predicted by the \citeauthor{Georgy} models with $0.3<\omega< 0.6$ \footnote{$\omega=\Omega_{in}/\Omega_{crit}$} at solar metallicity for stars in the corresponding region of the log\,\teff\ versus \logg\ diagram. 

As described in section \ref{cno}, we found that almost all stars in the sample show lower than solar C abundances and higher than solar N abundances. This indicates that the stars have undergone mixing and that the products of the H-burning CNO cycle are visible on the stellar surface. 
In the lower panel of Fig. \ref{logteff_logg}, the observed [N/C] abundance ratios of stars with 3.5 \Msol\ and 4.0 \Msol\ in the metallicity range $\mathrm{-0.1<[Fe/H]<0.1}$ are compared with the values predicted by the \citeauthor{Georgy} models for stars in the same region of the log\,\teff\ versus \logg\ diagram.  The stars are warmer than the maximum extent of the clump, or blue loop, where stars are located when they undergo central He-burning. However, because this phase is much longer than the crossing of the Hertzsprung gap, we consider it to be highly probable that all the stars are on the clump. 
We also note that the [N/C] abundance ratios for the majority of stars are consistent with the values predicted for clump stars in the \citeauthor{Georgy} models.
For all these quantities, a similar agreement is obtained with the models of \cite{Lagarde}, as shown in Fig. \ref{logteff_logg_g_l}.

\section{Conclusions}
We observed a sample of 26 bright giant stars that had a photometric metallicity estimated to be in the range $\mathrm{-2.5 \leq [Fe/H] \leq -1}$. 
The aim was a detailed inventory of the neutron-capture elements. The analysis of the sample showed that all stars were metal rich ($\mathrm{-0.4 \leq [Fe/H] \leq 0.4}$), with ages between 0.1 Gyr and 0.55 Gyr. Ten stars rotate rapidly ($\vseni >\ \mathrm{10\ km\, s^{-1}}$).
\\
The main conclusions of our study are listed below.
\begin{itemize}
    \item Photometric metallicity calibrations may show an age-metallicity degeneracy. 
   \item Several stars show radial velocity variability, with an error on the Gaia DR2 radial velocity above 1 \kms\ and/or a Gaia DR2 radial velocity that differs by more than 5$\sigma$ from our observed \vrad. Ten stars have also been identified as binary stars by \cite{Kervella} from proper motion anomalies. Stars HD 192045 and HD 213036 show radial velocity and proper motion variations. We therefore suggest that they are binary stars.
    \item The kinematic data indicate that all the stars are in prograde rotation around the Galactic centre on low-eccentricity orbits. This is typical of members of the thin-disc population.    
    \item Stellar masses between 2.5 \Msol\ and 6.0 \Msol\ suggest that the sample is composed of evolved stars that were of A and B type when they were on the main sequence. This hypothesis is supported by the chemistry of the stars, which is similar to that of A- and B-type stars in other studies.
    \item The A-type stars include chemically peculiar stars
    \citep[see the catalogue of][and references therein]{cpstars}. None of the stars observed by us shows any of the typical signs of CP stars, but they are similar to the normal A stars studied by \citet{Royer}. This suggests that chemical peculiarities are an atmospheric phenomenon that is erased when the stars evolve to the red giant phase.
    \item The derived \vseni\ agree with the theoretical values that have been predicted for these stars by two sets of stellar evolution models computed by the Geneva group with two different codes that include rotation effects in a similar way. This implies that the residual rotational velocity is the result of the secular expansion of the stellar radius in post-main-sequence evolution.     
    \item  The [C/Fe] ratio is lower than solar and [N/Fe] is higher than solar in all the stars except one. This suggests that the stars have undergone mixing and that the material shows the effects of CNO processing. This is in line with the predictions from stellar evolution models.
    \item The stars show a Ba enhancement but a low [s/Fe] ratio. This makes them similar to mild Ba stars. This might be due to a microturbulence that is higher than what we have assumed. 
    \item We did not robustly detect any correlation between chemical abundances and rotational velocities. Again, this is in line with theoretical predictions. 

\end{itemize}

\begin{acknowledgements}
We are grateful to the anonymous referee whose report helped us to improve our paper.
PB is grateful to Luca Casagrande for an interesting exchange on photometric metallicity calibrations.
We gratefully acknowledge support from the French National Research Agency (ANR) 
funded project ``Pristine'' (ANR-18-CE31-0017). 
This work was partially supported by the EU COST Action CA16117 (ChETEC).  
CC acknowledges support from the Swiss National Science Foundation (Project 200020-192039 PI CC). 
GM has received funding from the European Research Council (ERC) under the European Union's Horizon 2020 research and innovation programme (grant agreement No 833925, project STAREX).
This work has made use of data from the European Space Agency (ESA) mission
{\it Gaia} (\href{https://www.cosmos.esa.int/gaia}{https://www.cosmos.esa.int/gaia}), processed by the {\it Gaia}
Data Processing and Analysis Consortium (DPAC,
\href{https://www.cosmos.esa.int/web/gaia/dpac/consortium}{https://www.cosmos.esa.int/web/gaia/dpac/consortium}). Funding for the DPAC
has been provided by national institutions, in particular the institutions
participating in the {\it Gaia} Multilateral Agreement.

\end{acknowledgements}

%
%

\begin{appendix}

\section{Chemical abundances}

\begin{table}[ht!]
\caption{Elemental abundances of C, N, O with errors. }
\label{tab:cno}
\centering
\resizebox{\textwidth}{!}{
\begin{tabular}{lrrrrrrrrr}
\hline\hline 
Star & A(C) & $\sigma$ & [C/Fe] & A(N) & $\sigma$ & [N/Fe] & A(O) & $\sigma$ & [O/Fe] \\
 & dex & & dex & dex & & dex & dex & & dex \\
\hline
\\
HD 192045       & 7.90 & 0.30 & --0.49 & 8.31             & 0.40             & 0.56             & 8.72           & 0.12           & 0.23           \\
HD 191066       & 8.10 & 0.40 & --0.48 & 8.51             & 0.40             & 0.57             & 8.85           & 0.12           & 0.27           \\
HD 205732       & 8.20 & 0.30 & --0.39 & 8.52             & 0.40             & 0.57             & 8.72           & 0.10           & 0.18           \\
HD 213036       & 8.20 & 0.30 & --0.37 & 8.40             & 0.50             & 0.47             & \phantom{0.00} & \phantom{0.00} & \phantom{0.00} \\
HD 217089       & 8.00 & 0.20 & --0.42 & 8.25             & 0.25             & 0.47             & 8.72           & 0.10           & 0.01           \\
HD 9637         & 8.20 & 0.20 & --0.51 & 8.33             & 0.40             & 0.26              & \phantom{0.00} & \phantom{0.00} & \phantom{0.00} \\
HD 21269        & 7.80 & 0.20 & --0.75 & \phantom{0.00}   & \phantom{0.00}   & \phantom{0.00}   & \phantom{0.00} & \phantom{0.00} & \phantom{0.00} \\
HD 19267        & 7.95 & 0.20 & --0.47 & 8.18             & 0.40             & 0.40             & 8.71           & 0.09           & --0.26          \\
HD 13882        & 8.15 & 0.20 & --0.31 & 8.01             & 0.40             & 0.19             & \phantom{0.00} & \phantom{0.00} & \phantom{0.00} \\
HD 189879       & 8.00 & 0.50 & --0.51 & 8.45             & 0.50             & 0.58             & 8.78           & 0.07           & 0.04           \\
HD 195375       & 7.55 & 0.20 & --0.69 & 8.22             & 0.40             & 0.62             & 8.53           & 0.11           & --0.33          \\
HD 221232       & 8.00 & 0.20 & --0.50 & 8.34             & 0.40             & 0.48             & \phantom{0.00} & \phantom{0.00} & \phantom{0.00} \\
HD 219925       & 8.00 & 0.30 & --0.40 & 8.31             & 0.40             & 0.55             & 8.72           & 0.10           & 0.30           \\
HD 278          & 8.10 & 0.30 & --0.46 & 8.23             & 0.60             & 0.31             & \phantom{0.00} & \phantom{0.00} & \phantom{0.00} \\
HD 11519        & 8.15 & 0.20 & --0.39 & 8.33             & 0.40             & 0.43             & 8.78           & 0.08           & --0.02          \\
TYC 2813-1979-1 & 8.30 & 0.20 & --0.28 & 8.22             & 0.40             & 0.28             & 9.06           & 0.08           & 0.22           \\
BD+42 3220      & 8.80 & 0.30 & --0.10 & 8.37             & 0.40             & 0.11             & 8.91           & 0.21           & --0.26          \\
BD+44 3114      & 8.80 & 0.30 & --0.07 & 8.51             & 0.40             & 0.28             & \phantom{0.00} & \phantom{0.00} & \phantom{0.00} \\
TYC 3136-878-1  & 8.20 & 0.30 & --0.24 & 7.98             & 0.40             & 0.18             & 8.76           & 0.12           & 0.07           \\
HD 40509        & 8.15 & 0.30 & --0.47 & \phantom{0.00}   & \phantom{0.00}   & \phantom{0.00}   & \phantom{0.00} & \phantom{0.00} & \phantom{0.00} \\
HD 41710        & 8.00 & 0.20 & --0.48 & 8.33             & 0.40             & 0.49             & 8.71           & 0.07           & --0.22          \\
HD 40655        & 7.80 & 0.20 & --0.35 & 8.14             & 0.40             & 0.63             & 8.51           & 0.09           & --0.12          \\
HD 45879        & 7.25 & 0.20 & --0.89 & 7.34             & 0.40             & --0.16           & 8.45           & 0.07           & --0.38          \\
HD 55077        & 7.50 & 0.20 & --0.58 & 7.77             & 0.60             & 0.33             & \phantom{0.00} & \phantom{0.00} & \phantom{0.00} \\
HD 61107        & 8.05 & 0.20 & --0.50 & 8.26             & 0.60             & 0.35             & \phantom{0.00} & \phantom{0.00} & \phantom{0.00} \\
HD 63856        & 8.00 & 0.30 & --0.45 & 8.58             & 0.40             & 0.77             & 8.78           & 0.10           & 0.34           \\

\hline 
\end{tabular}
}
\end{table}

\begin{table*}
\caption{Elemental abundances of Mg, Al, and Ca with errors.}
\label{tab:mgalca}
\centering
\resizebox{\textwidth}{!}{
\begin{tabular}{lrrrrrrrrr}
\hline\hline 
Star & A(Mg) & $\sigma$ & [Mg/Fe] & A(Al) & $\sigma$ & [Al/Fe] & A(Ca) & $\sigma$ & [Ca/Fe] \\
 & dex & & dex & dex & & dex & dex & & dex \\
\hline
\\
HD 192045       & 7.71  & 0.09 &  0.28   &  6.52 & 0.01 &  0.16   & 6.24 & 0.10 & 0.01  \\
HD 191066       & 7.79  & 0.09 &  0.16   &  6.66 & 0.01 &  0.10   & 6.43 & 0.02 & 0.02  \\
HD 205732       & 7.93  & 0.10 &  0.30   &  6.72 & 0.07 &  0.15   & 6.48 & 0.15 & 0.05  \\
HD 213036       & 7.63  & 0.08 &  0.02   &  6.48 & 0.06 &  --0.05 & 6.37 & 0.10 & --0.02 \\
HD 217089       & 7.57  & 0.07 &  0.11   &  6.38 & 0.01 &  --0.01 & 6.29 & 0.06 & 0.04  \\
HD 9637         & 7.52  & 0.07 &  --0.22 &  6.50 & 0.01 &  --0.17 & 6.57 & 0.09 & 0.04  \\
HD 21269        & 7.70  & 0.36 &  0.11   &  6.22 & 0.11 &  --0.30 & 5.91 & 0.26 & --0.47 \\
HD 19267        & 7.57  & 0.08 &  0.12   &  6.35 & 0.06 &  --0.04 & 6.34 & 0.01 & 0.09  \\
HD 13882        & 7.60  & 0.08 &  0.10   &  6.30 & 0.06 &  --0.13 & 6.34 & 0.04 & 0.05  \\
HD 189879       & 7.69  & 0.09 &  0.14   &  6.56 & 0.07 &  0.08   & 6.38 & 0.15 & 0.03  \\
HD 195375       & 7.50  & 0.07 &  0.21   &  6.36 & 0.06 &  0.15   & 6.15 & 0.16 & 0.08  \\
HD 221232       & 7.56  & 0.07 &  0.02   &  6.41 & 0.06 &  --0.06 & 6.42 & 0.06 & 0.09  \\
HD 219925       & 7.71  & 0.09 &  0.27   &  6.43 & 0.06 &  0.07   & 6.27 & 0.10 & 0.05  \\
HD 278          & 7.32  & 0.22 &  --0.28 &  7.08 & 0.24 &  0.55   & 6.46 & 0.04 & 0.07  \\
HD 11519        & 7.65  & 0.08 &  0.07   &  6.39 & 0.06 &  --0.12 & 6.37 & 0.08 & 0.00  \\
TYC 2813-1979-1 & 7.80  & 0.10 &  0.18   &  6.37 & 0.06 &  --0.18 & 6.57 & 0.08 & 0.16  \\
BD+42 3220      & 7.94  & 0.01 &  0.00   &  6.74 & 0.06 &  --0.13 & 6.66 & 0.14 & --0.07 \\
BD+44 3114      & 8.04  & 0.09 &  0.12   &  6.74 & 0.10 &  --0.11 & 6.63 & 0.10 & --0.08 \\
TYC 3136-878-1  & 7.70  & 0.01 &  0.22   &  6.43 & 0.09 &  0.02   & 6.34 & 0.05 & 0.08  \\
HD 40509        & 7.41  & 0.24 &  --0.25 &  6.80 & 0.15 &  0.22   & 6.51 & 0.06 & 0.07  \\
HD 41710        & 7.80  & 0.19 &  0.29   &  6.31 & 0.05 &  --0.13 & 6.48 & 0.16 & 0.17  \\
HD 40655        & 7.20  & 0.14 &  0.01   &  6.21 & 0.05 &  0.10   & 6.09 & 0.13 & 0.11  \\
HD 45879        & 7.23  & 0.16 &  0.05   &  6.05 & 0.03 &  --0.07 & 6.27 & 0.07 & 0.29  \\
HD 55077        & 7.46  & 0.16 &  0.34   &  6.28 & 0.11 &  0.23   & 6.32 & 0.03 & 0.41  \\
HD 61107        & 7.60  & 0.17 &  0.01   &  6.50 & 0.10 &  --0.02 & 6.60 & 0.08 & 0.22  \\
HD 63856        & 7.70  & 0.01 &  0.20   &  6.56 & 0.07 &  0.14   & 6.33 & 0.06 & 0.05  \\

\hline 
\end{tabular}
}
\end{table*}

\begin{table*}
\caption{Elemental abundances of the n-capture elements Sr, Y, Ba, La, Ce, Pr, Nd, Sm, and Eu.} 
\label{tab:ncap}
\centering
\resizebox{\textwidth}{!}{
\begin{tabular}{lrrrrrrrrrr}
\hline\hline 
Star & [Sr/Fe] & [Y/Fe] & [Ba/Fe] & [La/Fe] & [Ce/Fe] & [Pr/Fe] & [Nd/Fe] & [Sm/Fe]  & [Eu/Fe] & [s/Fe] \\
 & dex & dex & dex & dex & dex & dex & dex & dex & dex & dex \\
\hline
\\
HD 192045        &         &  0.07    &  0.40  &  0.34   & 0.12   & --0.37  & 0.43   & 0.27   & 0.25    & 0.23   \\
HD 191066        &  0.16   &  0.22    &  0.61  &  0.34   & 0.22   & --0.36  & 0.38   & 0.38   & 0.31    & 0.29   \\
HD 205732        &  0.20   &  0.11    &  0.35  &  0.28   & 0.26   &         & 0.32   & 0.27   & 0.40    & 0.24   \\
HD 213036        &         &  0.18    &  0.82  &  0.25   & 0.18   & --0.38  & 0.24   & 0.19   & 0.32    & 0.21   \\
HD 217089        &  0.03   &  --0.01  &  0.68  &  0.01   & 0.04   & --0.68  & 0.05   & 0.00   & 0.08    & 0.02   \\
HD 9637          &  --0.11 &  --0.20  &  0.54  &  --0.08 & --0.05 &         & --0.14 &        & 0.04    & --0.12 \\
HD 21269         &  0.04   &  0.25    &  1.79  &  --0.18 & 0.15   &         & 0.01   &        & 0.09    & 0.06   \\
HD 19267         &  --0.22 &  --0.32  &  0.42  &  --0.30 & --0.27 & --1.23  & --0.16 & --0.21 & --0.12  & --0.26 \\
HD 13882         &  --0.16 &  --0.15  &  0.59  &  --0.28 & --0.25 & --1.26  & --0.14 & --0.24 & --0.11  & --0.21 \\
HD 189879        &         &  --0.09  &  0.55  &  0.03   & --0.04 & --0.77  & 0.02   & --0.03 & 0.05    & --0.02 \\
HD 195375        &  --0.22 &  --0.31  &  0.43  &  --0.29 & --0.26 & --1.12  & --0.20 & --0.30 & --0.12  & --0.26 \\
HD 221232        &  --0.07 &  --0.26  &  0.78  &  --0.24 & --0.16 &         & --0.15 & --0.25 & --0.12  & --0.20 \\
HD 219925        &  0.43   &  0.19    &  0.68  &  0.26   & 0.24   & --0.18  & 0.35   & 0.30   & 0.38    & 0.26   \\
HD 278           &  --0.27 &  --0.01  &  0.48  &  0.06   & --0.06 &         & 0.00   &        & 0.03    & 0.00   \\
HD 11519         &  0.14   &  --0.05  &  0.69  &  --0.08 & --0.15 & --0.94  & 0.01   & 0.01   & --0.06  & --0.07 \\
TYC 2813-1979-1  &  0.00   &  0.01    &  0.55  &  0.03   & --0.04 & --0.87  & 0.12   & 0.22   & 0.01    & 0.03   \\
BD+42 3220       &  --0.03 &  0.03    &  0.02  &  --0.25 & --0.42 & --1.58  & --0.21 & 0.09   & --0.08  & --0.21 \\
BD+44 3114       &  --0.10 &  0.01    &  0.20  &  --0.22 & --0.59 & --1.73  & --0.13 & 0.12   & 0.00    & --0.23 \\
TYC 3136-878-1   &  0.05   &  0.11    &  0.40  &  0.08   & --0.04 & --0.73  & 0.22   & 0.27   & 0.10    & 0.09   \\
HD 40509         &  --0.14 &  --0.18  &  0.71  &  --0.31 & --0.43 &         & --0.27 &        & --0.24  & --0.30 \\
HD 41710         &  0.11   &  0.02    &  1.46  &  --0.11 & --0.08 & --1.01  & 0.08   & --0.22 & --0.04  & --0.02 \\
HD 40655         &  0.11   &  --0.18  &  1.06  &  --0.21 & --0.13 & --0.76  & --0.07 & --0.27 & 0.01    & --0.15 \\
HD 45879         &  --0.19 &  --0.08  &  1.46  &  --0.31 & --0.08 &         & --0.12 & --0.27 & --0.19  & --0.15 \\
HD 55077         &  0.03   &  --0.21  &  0.88  &  0.01   & 0.04   &         & 0.00   &        & 0.13    & --0.04 \\
HD 61107         &  --0.19 &  --0.18  &  0.86  &  --0.11 & --0.18 &         & --0.02 &        & --0.04  & --0.12 \\
HD 63856         &  0.10   &  0.36    &  0.55  &  0.38   & 0.31   & --0.14  & 0.47   & 0.21   & 0.40    & 0.38   \\
\hline  
\end{tabular}
}
\end{table*}

\begin{table*}
\caption{Estimated errors in element abundance ratios [X/Fe] for neutron-capture elements for the star HD 13882.} \label{errors}
\centering
\begin{tabular}{l r r r }
\hline\hline
[X/Fe]            &   $\Delta \teff$ =   & $\Delta$ \logg =   &$\Delta$  $\xi$ =   \\     
                    &       100~K                 &    0.5~dex        &     0.5~\kms \\  
\hline
Sr &   0.10  &  0.15  &  --0.25  \\
Y  &   0.10  &  0.10  &  --0.25  \\
La &   0.10  &  0.20  &  --0.10  \\  
Ce &   0.05  &  0.05  &  --0.07  \\
Pr &   0.05  &  0.15  &  --0.05  \\
Nd &   0.15  &  0.20  &  --0.05  \\
Sm &   0.05  &  0.20  &  --0.08  \\
Ba &   0.20  &  0.25  &  --0.10  \\
Eu &   0.05  &  0.20  &  --0.05  \\

\hline
\end{tabular}
\end{table*}

\end{appendix}

\end{document}